# The role of hydrophobic interactions in folding of β-sheets

Jiacheng Li[a,1], Xiaoliang Ma[a,1], Hongchi Zhang[a,1], Chengyu Hou[b,1], Liping Shi[a], Shuai Guo[a], Chenchen Liao[b], Bing Zheng[c], Lin Ye[d], Lin Yang[a,d,*], Xiaodong He[a,e]

[a] *National Key Laboratory of Science and Technology on Advanced Composites in Special Environments, Center for Composite Materials and Structures, Harbin Institute of Technology, Harbin 150080, China*

[b] *School of Electronics and Information Engineering, Harbin Institute of Technology, Harbin 150080, China*

[c] *Key Laboratory of Functional Inorganic Material Chemistry (Ministry of Education) and School of Chemistry and Materials Science, Heilongjiang University, Harbin 150001, P. R. China.*

[d] *School of Aerospace, Mechanical and Mechatronic Engineering, The University of Sydney, NSW 2006, Australia*

[e] *Shenzhen STRONG Advanced Materials Research Institute Co., Ltd, Shenzhen 518035, P. R. China.*

**Exploring the protein-folding problem has been a long-standing challenge in molecular biology. Protein folding is highly dependent on folding of secondary structures as the way to pave a native folding pathway. Here, we demonstrate that a feature of a large hydrophobic surface area covering most side-chains on one side or the other side of adjacent β-strands of a β-sheet is prevail in almost all experimentally determined β-sheets, indicating that folding of β-sheets is most likely triggered by multistage hydrophobic interactions among neighbored side-chains of unfolded polypeptides, enable β-sheets fold reproducibly following explicit physical folding codes in aqueous environments. β-turns often contain five types of residues characterized with relatively small exposed hydrophobic proportions of their side-chains, that is explained as these residues can block hydrophobic effect among neighbored side-chains in sequence. Temperature dependence of the folding of β-sheet is thus attributed to temperature dependence of the strength of the hydrophobicity. The hydrophobic-effect-based mechanism responsible for β-sheets folding is verified by bioinformatics analyses of thousands of results available from experiments. The folding codes in amino acid sequence that dictate formation of a β-hairpin can be deciphered through evaluating hydrophobic interaction among side-chains of an unfolded polypeptide from a β-strand-like thermodynamic metastable state.**

## INTRODUCTION

Protein products are the basis of life on Earth and serve nearly all the functions in the essential biochemistry of life science. Each nascent protein exists as an unfolded polypeptide or random coil when translated from a sequence of mRNA to a linear chain of residues by a ribosome. The intrinsic biological functions of a protein are expressed and determined by its native three-dimensional (3D) structure that derives from the physical process of protein folding[1], by which a polypeptide folds into its native characteristic and functional three-dimensional structure, in an expeditious and reproducible manner. Protein folding can thereby be considered the most important mechanism, principle, and motivation of biological existence, functionalization, diversity, and evolution[2-4].

Based on the complexity of protein folding, the protein-folding problem has been summarized in three unanswered questions[1]: (i) What is the physical folding code in the amino acid sequence that dictates the particular native 3D structure? (ii) What is the folding mechanism that enables proteins to fold so quickly? (iii) Is it possible to devise a computer algorithm to effectively predict a protein's native structure from its amino acid sequence? Moreover, another essential question is why protein folding highly depends on the solvent (water or lipid bilayer)[5] and the temperature[6]? The protein folding problem was brought to light over 60 years ago. In particular, since Anfinsen shared a 1972 Nobel Prize in Chemistry for his work revealing the connection between the amino acid sequence and the native conformation[7], understanding of protein sequence-structure relationships has become the most fundamental task in molecular and structural biology [8].

Protein folding is one of the miracles of nature that human technology finds quite difficult to follow, due to the very large number of degrees of rotational freedom in an unfolded polypeptide chain. In the 1960s, Cyrus Levinthal pointed out that the apparent contradiction between the astronomical number of possible conformations for a protein chain and the fact that proteins can fold quickly into their native structures should be regarded as a paradox (Levinthal's paradox) [9], so there must be mechanisms that allow polypeptide chains to find the native states encoded in their sequence. As stated in Anfinsen's Dogma, the well-defined native 3D structures of small globular proteins are uniquely encoded in their primary structures (the amino acid sequences), is kinetically reproducible and stable under a range of physiological conditions, and can therefore be considered as an issue of the certainty.

Many proteins or protein domains, relatively rapid and efficient refolding can be observed in vitro, thus proteins may be regarded as "folding themselves" following explicit folding pathways[1]. Protein folding is considered a free energy minimization or a relaxation process that is guided mainly by the following physical forces: (i) formation of intramolecular hydrogen bonds, (ii) van der Waals interactions, (iii) electrostatic interactions, (iv) hydrophobic interactions, (v) chain entropy of protein, (vi) thermal motions[1,10]. Among them, hydrophobic effect is normally thought to play a decisive role[11]. Currently, the generally accepted hypothesis in the field is to conceive of protein folding in a funnel-shaped energy landscape, where every possible conformation is represented by a free energy value. The rapid folding of proteins has been attributed to random thermal motions that cause conformational changes leading energetically downhill toward the native structure corresponds to its free energy minimum under the solution conditions [1,10]. However, there are both enthalpic and entropic contributions to free energy of protein that change with temperature and so give rise to heat denaturation and, in some cases, cold denaturation[12]. So far the hypothesis haven't been able to decipher the folding code and therefore aren't generally able to read a sequence and predict what shape it will adopt.

The interaction of protein surface with the surrounding water is often referred to as protein hydration layer (also sometimes called hydration shell) and is fundamental to structural stability of protein, because non-aqueous solvents in general denature proteins [13]. The hydration layer around a protein has been found to have dynamics distinct from the bulk water to a distance of 1 nm and water molecules slow down greatly when they encounter a protein[14]. Thus, hydrophilic side chains of proteins are normally hydrogen bonded with surrounding water molecules in aqueous environments, thereby preventing the surface hydrophilic side-chains of proteins from randomly hydrogen bonding together [14,15 16]. This is the reason why proteins usually do not aggregate or crystallize in unsaturated aqueous solutions[17] , even though the solvent-facing surface of the proteins is usually composed of predominantly hydrophilic regions. Experiments have also shown that secondary structures of protein (such as α-helices and β-sheets) are stabilized by hydrogen bonds between the N-H groups and C=O groups of the main chain[18,19]. This also indicates that the shielding effect of surrounding water molecules prevent hydrophilic side-chains from interfering with the formation of secondary structures during protein folding. Thus, water molecules should be able to saturate the hydrogen bond formations of hydrophilic side-chains and the main chain before the protein folding [14-16], due to water molecules have very strong polarity[20,21].

This is the reason why intrinsically disordered proteins (IDPs) and regions (IDRs) can make up a significant part of the proteome [22]. Before the folding of secondary structures, the early steps of protein folding may be not directly dominated by the formation of intramolecular hydrogen bonds, due to the shielding effect of surrounding water molecules. Thus, this problem may lie in our lack of understanding of the hydrophobic interaction among neighbored side-chains of unfolded proteins at early steps of the folding, given the lack of awareness of the importance of the shielding effect of water.

Almost all experimentally determined native tertiary structures of water-soluble proteins have a hydrophobic core in which hydrophobic side-chains are buried from water[23-25]. Incidentally, polar residues interact favorably with water, thus the solvent-facing surface of the peptide is usually composed of predominantly hydrophilic regions[26]. Minimizing the number of hydrophobic side-chains exposed to water, namely, hydrophobic collapse thus has been regarded as one of the most important driving force for protein folding processs[27]. Experimental methods such as laser temperature jumping technology and single molecule experimental techniques have revealed that protein folding first leads to the formation of secondary structures (α-helices and β-strands), and the tertiary structure is formed by the folding of secondary structures [28]. It is likely that the nascent polypeptide forms initial secondary structure through creating localized regions of predominantly hydrophobic residues due to hydrophobic effect[29]. The secondary structures interacts with water, thus placing thermodynamic pressures on these regions which then aggregate or "collapse" into a tertiary conformation with a hydrophobic core[26]. Therefore, protein folding is highly dependent on folding of secondary structures as the way to hierarchically pave a native folding pathway that lead to formation of correct tertiary structures and cause conformational changes leading energetically downhill toward the native globular structure that possesses the minimum free energy. Thus, decipher of the folding codes in amino acid sequence that dictate the secondary structures formation should be regarded as a key to crack the protein folding problem. Among types of secondary structure in proteins, the β-sheet is the most prevalent. If the controlling mechanism for β-sheet folding can be revealed, it would remarkably promote solution of the protein folding problem.

Currently, several hypotheses has been proposed for explaining the folding mechanism of β-sheet. The hydrophobic zipper hypothesis indicates that a hairpin is first formed before hydrophobic contacts act as constraints which bring other contacts into spatial proximity[30]. This leads to further constrain and causes the rest of the contacts to zip up. Munoz *et al* proposed that the folding of a *β*-hairpin initiates at the turn and propagates towards the tails[31]. In particular, they found that stabilization through hydrophobic contacts between residues and hydrogen bonding interaction are important for the formation of the *β*-hairpin. Petrovich *et al.* [32] studied a 37-residue triple-stranded *β*-sheet protein via MD simulations. Their results indicate that a β-hairpin first appears before the third strand joins in to complete the β-sheet at the end of the folding process. They ascribe the folding mechanism of the β-sheet to a combination of initial hydrophobic collapse and zipper mechanism, which serve to nucleate the hairpin formation. Notably, all the three mechanisms above suggest that the folding of a β-sheet is necessarily preceded by the occurrence of a β-turn. We are still missing a "folding mechanism" for β-sheets. By mechanism, we mean a narrative that explains how the time evolution of a β-sheet folding development derives from its amino acid sequence and solution conditions.

**Results**

β-sheet folding highly depends on the temperature [5], where β-sheets can form in as little as 1 microsecond after the temperature jumping[33-35]. β-sheets consist of β-strands connected laterally by at least three backbone hydrogen bonds, forming a generally pleated sheet. A β-strand is a stretch of polypeptide chain typically 3 or more amino acids long with backbone in an extended conformation. It most like that the β-strands exist before the folding of β-sheets. Because it is difficult to explain how the folding process of a β-sheet (i.e., laterally hydrogen bonding process of segments of unfolded polypeptide) is accompanied by stretching process of the segments of polypeptide into β-strands. There must be mechanisms that allow polypeptide chain segments to find the states of β-strands encoded in their sequence. There also must be some physical effects providing the long-range attractive force among β-strands for the β-sheets formation.

Experimental evidences of the folding of unfolded proteins provide corroboration for a hypothesis that folding initiation sites arise from hydrophobic interactions [11,36]. The folding of β-strands and β-sheets may be driven by hydrophobic interactions, as the nascent polypeptide may form initial primary structure through creating localized regions of predominantly hydrophobic residues[29]. Hydrophobic effect most likely can contribute to the formation of β-sheets through multistage aggregations of neighbored hydrophobic groups of unfolded polypeptides, which lead to the formation of β-strands, and consequently fold into β-sheets. A β-sheet always is amphipathic in nature, namely, contain hydrophilic surface areas and hydrophobic surface areas. Note that the hydrophobic attraction (due to the hydrophobic effect) among adjacent side-chains on one side or the other side of a β-strand may be common in experimentally determined protein structures, which should be considered as an evidence for hydrophobic effect dominating the formation of β-strands.

It has previously been noted that many amino acid side chains contain considerable nonpolar sections, even if they also contain polar or charged groups[36]. Namely, hydrophilic side-chains are not completely hydrophilic. The hydrophilicity of hydrophilic side-chains is normally expressed by C=O or N-H2 groups at their ends, and the other portions of hydrophilic side-chains are hydrophobic, because the molecular structures of these portions are basically alkyl and benzene ring structures, as shown in Figure 1. Folding initiation sites of β-brands might therefore contain not only accepted “hydrophobic” amino acids, but also larger hydrophilic side-chains[36]. If formation of β-brands is driven by hydrophobic interactions among neighbored side-chains of unfolded polypeptide, we should be able to find experimental evidence of the hydrophobic interaction in the Protein Data Bank (PDB) achieves, due to hundreds of thousands of β-sheet structures have been experimentally determined. In an aqueous environment, the water molecules tend to segregate around the “hydrophobic” side chains of the nascent protein, creating hydration shells of ordered water molecules[37]. An ordering of water molecules around a hydrophobic region increases order in a system and therefore contributes a negative change in entropy (less entropy in the system)[38]. The water molecules are fixed in these water cages which drives the hydrophobic collapse, or the aggregation of the hydrophobic groups. Thus, the hydrophobic interaction among neighbored side-chains in sequence can introduce entropy back to the system via the breaking of their water cages which frees the ordered water molecules[39]. If hydrophobic interactions among neighbor side-chains in amino acid sequences provide the structural stability for β-brands formation, we must can find out that the phenomenon of a large hydrophobic surface area covering on one side or the other side of a β-strand is prevail in almost all experimentally determined β-sheets. If the phenomenon of hydrophobic side-chains tend to cluster together on one side of adjacent β-strands of a β-sheet is prevail in almost all experimentally determined β-sheets, we may demonstrate that the hydrophobic interaction among the neighbored side-chains responsible for β-sheet-folding initiation.

The capability of an amino acid residue to get involved in the hydrophobic attraction with neighbored residues in sequence can be evaluated by the exposed alkyl and benzene ring structures of the side-chain, as shown in Fig.1, in which 20 kinds of amino acid residue are divided into four groups[40]. Arginine-R, Histidine-H, and Lysine-K can involve in hydrophobic interaction with adjacent hydrophobic side-chains in sequence due to their long hydrophilic side chains contain long nonpolar alkyl structures, see Fig.1A. Cysteine-C, Isoleucine-I, Leucine-L, Methionine-M, Tryptophan-W, Phenylalanine-F, Tyrosine-Y, and Valine-V can fully involve in hydrophobic interaction with adjacent side-chains due to their high hydrophobicity, see Fig.1B. Glutamate-E, Glutamine-Q, Threonine-T, and Alanine-A would allow limited participation in hydrophobic interaction with neighbored side-chains in sequence due to their exposed hydrophobic proportions is relatively small, see Fig.1C. Aspartate-D, Asparagine-N, Serine-S, Proline-P, and Glycine-G basically can't participate in hydrophobic interaction with adjacent side-chains in sequence due to the hydrophobic proportions of their side-chains are too small or being occluded by hydrophilic groups, see Fig.1D.

A de novo designed protein with curved β-sheet (PBDID: 5TPJ) is a good example for illustrating the phenomenon of the hydrophobic attraction (due to the hydrophobic effect) among adjacent side-chains on one side of each β-strand of the protein, see Fig.2[41]. To illustrate the hydrophobic attraction, we highlight the hydrophobic surface areas of adjacent side-chains on each β-strand of the protein based on the experimentally determined protein structure as shown in Fig. 2C and 2D. Noting that every β-strand is characterized by a large hydrophobic surface fully covering one side of the β-brand (the inner side), and caused each side-chains is parallel to every other side-chain of each strands due to the hydrophobic interaction. Parallel distribution of adjacent peptide planes of these β-strands also causes adjacent side-chains to distribute on opposite sides of the main chain and each carbonyl oxygen atom in a peptide plane tends to hydrogen bond with an amide hydrogen atom in an adjacent peptide plane due to the electrostatic attractions between them, except the Proline-P[15]. Parallel distribution of neighbored "hydrophobic" side-chains in a β-strand can effectively introduce entropy back to the system via the merging of the water cages of the side-chains which frees the ordered water molecules, see Fig.2D. Thus, β-strand should be considered as a metastable state for unfolded polypeptides corresponds to its free energy minimum under the solution conditions, creating localized regions of predominantly hydrophobic side-chains[15].

We use another small-molecule protein (PBDID:1OUR) as the example to demonstrate the role of hydrophobic interactions among neighbored side-chains played in formation of β-strands, β-turns and β-sheets, see Fig.3. The protein is mainly composed with β-strands and 10 β-turns. Every β-strand of the protein is also characterized by a large hydrophobic surface fully covering one side or the other side of the β-brand, see Fig.3A. Aspartate-D, Asparagine-N, Serine-S, Proline-P, Glycine-G most likely contribute to formation of β-turns in protein folding, due to the other neighbored side-chains in amino acid sequence tend to hydrophobic attract with each other through bypassing these residues (see Fig.1d). Thereby, Aspartate-D, Asparagine-N, Serine-S, Proline-P, Glycine-G can be classified as a hydrophobic blocking ($R_B$) group. It is worth noting that almost all the 10 β-turns of the protein are composed with two or more residues of Aspartate-D, Asparagine-N, Serine-S, Proline-P, Glycine-G, see Fig.3A and 3B. This indicates that two or more adjacent $R_B$ residues can effectively block hydrophobic attraction among neighbored side-chains in sequence at both side of a strand. We plot the protein structure into three parts according to three segments of the amino acid sequence to illustrate the hydrophobic collapse among neighbored β-strands in sequence, see Fig.3B and 3C. Hydrophobic interactions among these β-strands may drive them collapse together through bending the unfolded polypeptide at the location of these $R_B$ residues, namely, bypassing these $R_B$ residues at the turns to achieve the hydrophobic collapse. This also indicates that hydrophobic attraction among neighbored side-chains drive the β-strands formation and then cause hydrophobic attraction among the neighbored β-strands and formation of the β-sheets, due to β-strands formation create localized regions of predominantly hydrophobic residues and place thermodynamic pressures on these regions under the solution conditions. Formation of β-sheets also make β-strands aggregate or "collapse" into a tertiary conformation with a hydrophobic core. Thereby, we speculate that folding of β-sheets is triggered by multistage hydrophobic interactions among neighbored side-chains of unfolded polypeptides, enable β-sheets fold reproducibly following explicit physical folding codes in aqueous environments.

We use 1000 experimentally determined small protein structures to further demonstrate the hydrophobic-effect-based folding mechanism for β-sheets. All the 1000 small proteins were randomly selected from the PDB. 3235 β-strands can be identified in the 1000 protein structures by using the PDB archive and the STRIDE software[42]. From analysis of all the 3235 β-strands of the 1000 proteins in PDB, we find out that the feature of hydrophobic attraction (due to the hydrophobic effect) among adjacent side-chains on one side or the other side of a β-strand covering the length of the β-strand is prevail in all the experimentally determined β-strands (see Supplementary S5). This indicates that the hydrophobic interaction among the neighbored side-chains responsible for the formation of β-strands.

Aspartate-D, Asparagine-N, Serine-S, Proline-P, Glycine-G can't effectively hydrophobic attract with neighbored side-chains in sequence, see Fig.1D. Thus, Aspartate-D, Asparagine-N, Serine-S, Proline-P, Glycine-G most likely lead to β-turns formation in protein folding, due to the other neighbored side-chains in amino acid sequence tend to hydrophobic attract with each other through bypassing these residues. The β-turn is the third most important secondary structure after helices and β-strands. β-turns have been classified according to the values of the dihedral angles φ and ψ of the central residue. β-turns can be easily identified in between β-strands or α-helices of the protein structures using the PDB archive and the STRIDE software[42]. We identified 5776 β-turns in the 1000 protein structures, include about 1780 β-hairpin turns. We find out that about 97.4% of the β-turns contain at least one Aspartate-D, Asparagine-N, Serine-S, Proline-P or Glycine-G residue[43], as illustrated in Supplementary 2. Whereas, most of the rest no-$R_B$ β-turns contain at least one Glutamate-E, Glutamine-Q, Threonine-T, and Alanine-A residue. This indicates that Glutamate-E, Glutamine-Q, Threonine-T, and Alanine-A may contribute to the formation of β-turns due to their exposed hydrophobic proportions is relatively small. Moreover, about 99.3% β-hairpin turns contain at least one Aspartate-D, Asparagine-N, Serine-S, Proline-P or Glycine-G residue, see Supplementary 2.

Two $R_B$ residues coded together normally shouldn't be able to present at the middle of a long straight β-strand. Because the other residues of the strand at both sides of the two $R_B$ residues tend to hydrophobic aggregate together and thus would bend the strand at the two $R_B$ residues to achieve the hydrophobic interaction. However, we can still identified 29 long β-strands (each β-strands contain more than 12 residues), which are characterized by two adjacent $R_B$ residues locating at the middle of the β-strands through scanning the 1000 protein structures by using the STRIDE software[42]. By checking these long β-strands using PyMOL software, we find out that 24 of these long β-strands actually curved exactly at their two $R_B$ residues in the amino acid sequences, demonstrating the capability of $R_B$ residues to cause β-turns formation, see Fig. 4. The other 5 long β-strands either have three or more $R_B$ residues coded together or have $R_B$ residues located at one end of the strands that make the hydrophobic blocking region extend to the ends of these β-strands, thus undermining the hydrophobic interaction between the both ends of these β-strands, see Supplementary S3. The long β-strand of the 1YV7 protein curved at a sequence segment of threonine-threonine-terine-glutamate (TTSE), see Supplementary S3. This indicates that Glutamate-E, Glutamine-Q, Threonine-T, and Alanine-A may also contribute to the formation of β-turns due to their exposed hydrophobic proportions is relatively small.

The spike (S) protein of novel severe acute respiratory syndrome coronavirus 2 (SARS-CoV-2) is of great concern due to the coronavirus disease 2019 (COVID-19) pandemic. The D614G mutation in SARS-CoV-2 begin to receive widespread attention for its rising dominance worldwide. This mutation changes the amino acid at position 614, from D (aspartic acid) to G (glycine), the initial D614 is now the G614 variant. It is worth noting that the amino acid at position 614 is located at a β-turn in a tertiary structure of the spike. This is consistent with our new theory that both D (aspartic acid) to G (glycine) can result in the β-turn formation. The D-614-G nutation may accelerate the folding of the quaternary structure of the spike due to G614 most likely can contribute to the hydrophobic effect between two tertiary structures of the protein rather than the D614 (see Fig.1D), due to the position 614 located at the docking site in between them.

A typical β-hairpin structure contains two β-strands with hydrophobic attraction between each side-chain and every other side-chain on the strands. Thus, we might be able to predict β-hairpin structures through evaluating hydrophobic attraction among each side-chain with every other side-chain in the primary structure of a protein. We may can predict β-hairpin through identifying two neighbored sequences of residues in the polypeptide chain both characterized by hydrophobic attraction between each side-chain to every other side-chain, and have two $R_B$ in between them. By using this method, we identified 553 samples in terms of the characteristics above from the 1000 proteins. We find that 158 of the samples are β-hairpins, 36 of the samples are structures of strand-turn-strand, 296 of the samples are structures of strand-turn-helix, 23 of the samples are structures of coil-turn-strand, 23 of the samples are coil-turn-coil and 6 of the samples are α-helices. Thus, physical folding codes for β-hairpins and strand-turn-strand can be deciphered through evaluating hydrophobic interaction among side-chains of an unfolded polypeptide. The results show that strand-turn-helix also can be predict by the method. This indicates that folding of α-helix may be initiated from a β-strand-like thermodynamic metastable state[15].

## Conclusion

Many amino acid residues contain considerable nonpolar sections in their side-chains, even if they also contain polar or charged groups. This make hydrophobic interaction among neighbored amino acid side-chains in amino acid sequence of polypeptides becomes an important driving force for the stabilization of initial thermodynamic state of unfolded Proteins. The feature of a large hydrophobic surface area covering most side-chains on one side or the other side of adjacent β-strands of a β-sheet is prevail in almost all experimentally determined β-sheets. Minimizing the exposed hydrophobic portions of adjacent side-chains to water should be regarded as the most important driving force for the β-strands formation and caused each side-chains is parallel to every other side-chain on strands. β-turns often contain residues of Aspartate-D, Asparagine-N, Serine-S, Proline-P, Glycine-G which characterized with their side-chains having very small hydrophobic proportions exposure, that is explained as these residues can block hydrophobic effect among neighbored side-chains in sequence, thereby contribute to turns formation. The folding of β-sheets are most likely triggered by multistage hydrophobic interactions among neighbored side-chains of unfolded polypeptides, enable β-sheets fold reproducibly following explicit physical folding codes in aqueous environments. Temperature dependence of the folding of β-sheet is thus attributed to temperature dependence of the strength of the hydrophobicity. The hydrophobic collapse of β-strands into β-sheets most likely trigger enthalpy-entropy compensation of unfolded polypeptides, enable the main-chain of β-strands to get rid of the hydrogen-bonded water molecules and laterally hydrogen bonding with each other. The folding codes in amino acid sequence that dictate the formation of a β-hairpin can thus be deciphered through evaluating hydrophobic interaction among side-chains of an unfolded polypeptide from a β-strand-like thermodynamic metastable state.

# Materials and Methods

## Protein structures

In this study, many experimentally determined native structures of proteins are used to study the folding mechanism of β-sheets. All the three-dimensional (3D) structure data of protein molecules are resourced from the PDB database. IDs of these proteins according to PDB database are marked in the Fig.2, Fig.3, and Fig.4. In order to show the distribution of hydrophobic areas on the surface of β-strands and β-sheets in these figures, we used the structural biology visualization software PyMOL to display the hydrophobic surface areas of these secondary structures.

## Identification of secondary structures of proteins

Secondary structures of β-strands, β-turns, β-sheets and α-helices were identified in the 1000 proteins by using the STRIDE software[42]. We also used molecular 3D structure display software PyMOL to confirm the identification of secondary structures of proteins.

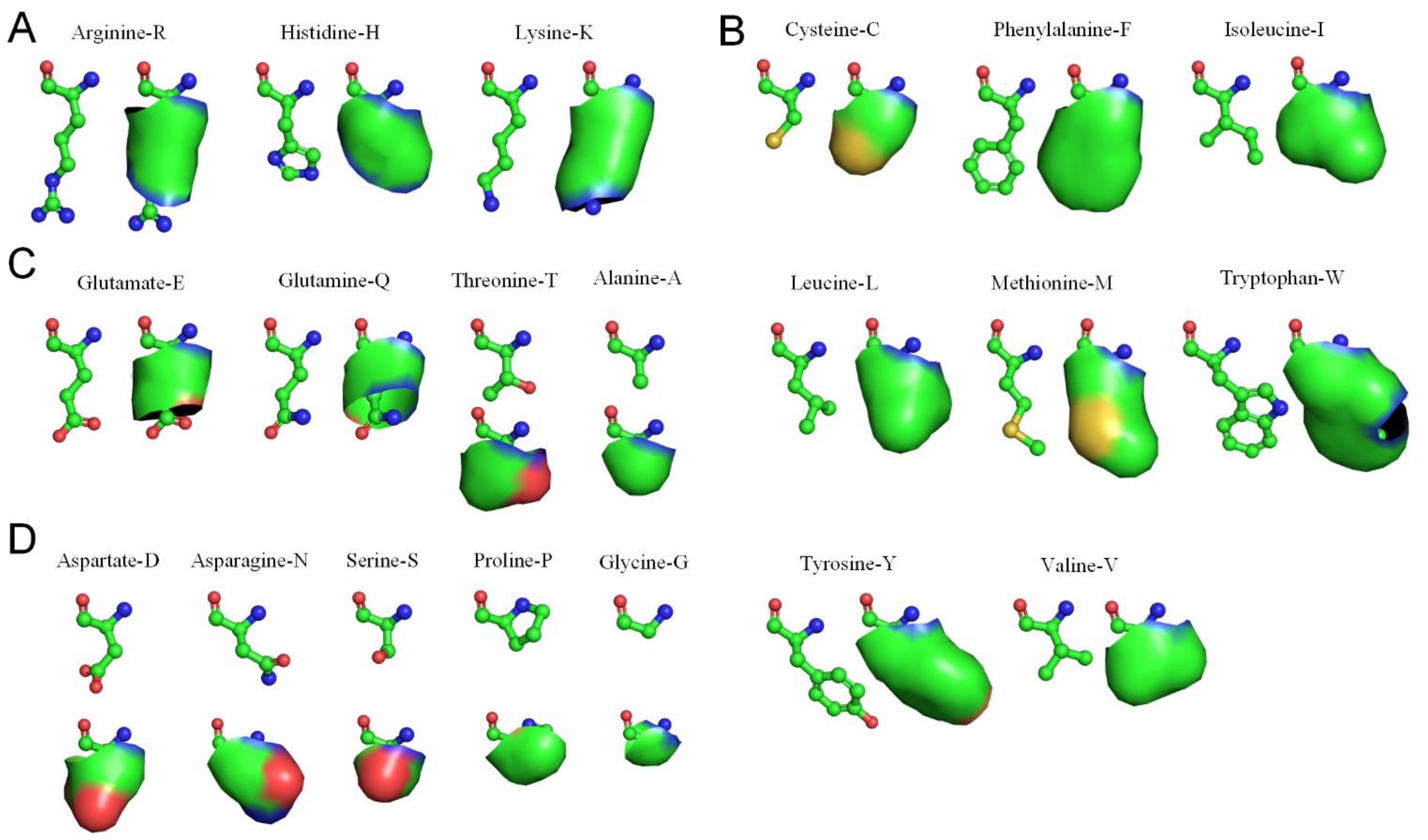

Figure 1 Hydrophobic portions of amino acid side-chains (hydrophobic portions are highlighted by green)

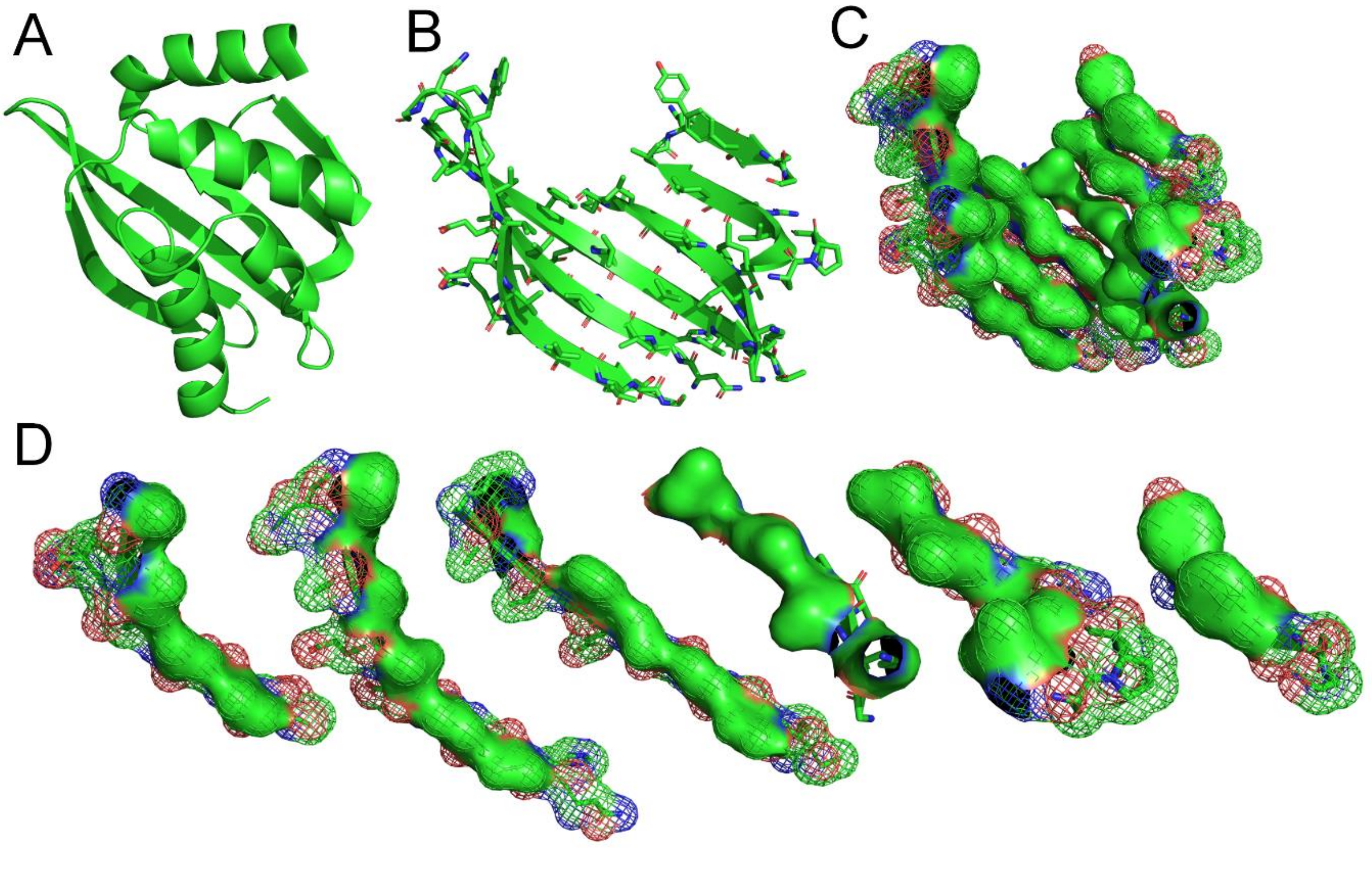

**Fig. 2. Hydrophobic attraction among neighbored side-chains of β-strands.** (**A**) A de novo designed protein (PBDID: 5TPJ). (**B**) The curved β-sheet of 5TPJ. (C) Hydrophobic attraction among adjacent β-strands via hydrophobic surface of side-chains of the β-sheet (hydrophobic surface is highlighted by using green surface areas). (D) Hydrophobic surface areas on the 6 β-strands of the sheet (green surface areas).

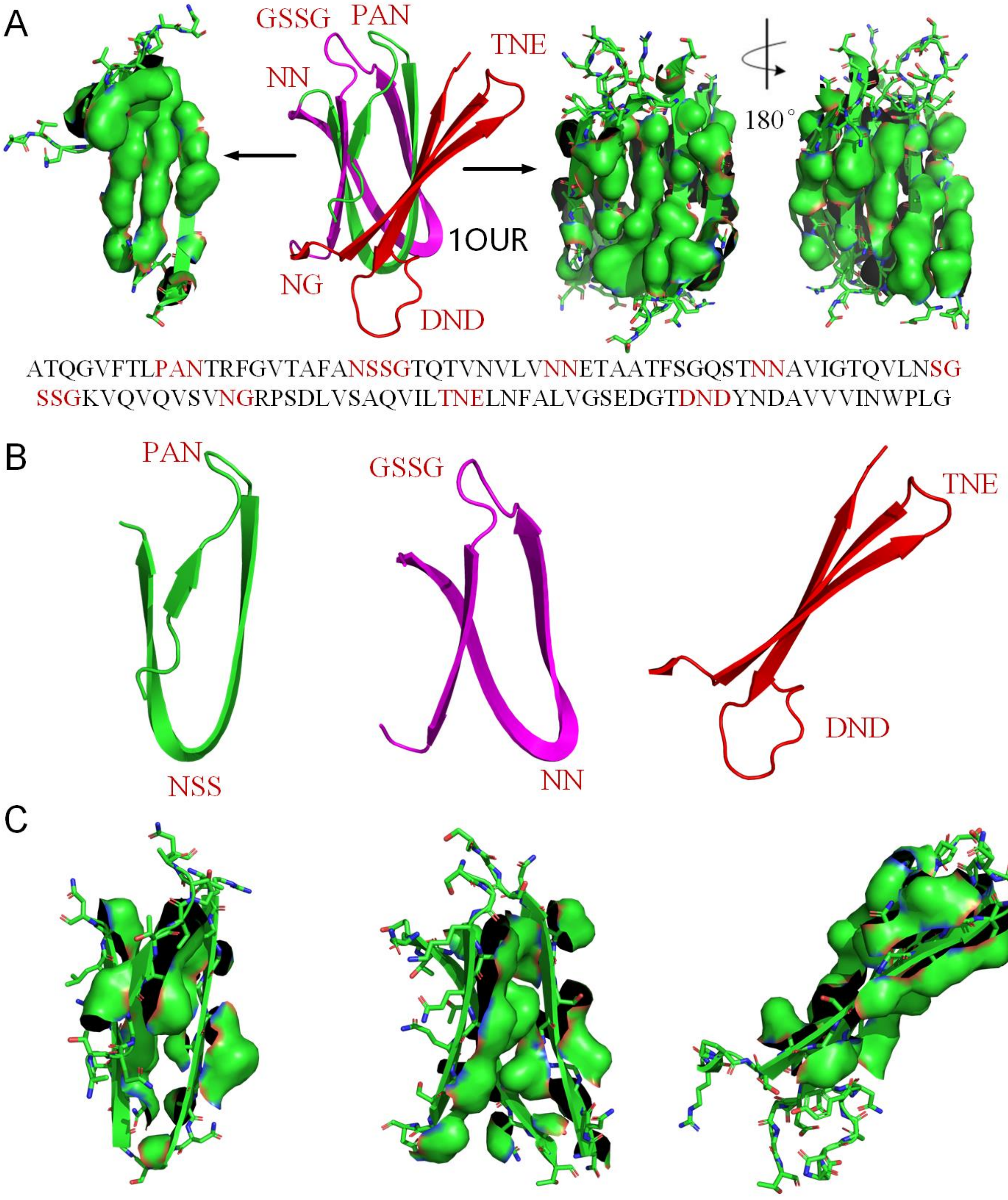

**Fig. 3.** (**A**) Hydrophobic surface areas on the β-strands of the protein 1OUR (hydrophobic surface of side-chains is highlighted by using green surface areas), residues located at turns are highlighted in red color in the sequence of the protein. (B) The parts of the protein (residues1-33 are highlighted in green, residues 34-71 are highlighted in magenta, residues 72-114 are highlighted in red). Hydrophobic surface areas on the β-strands of the sheet (green surface areas).

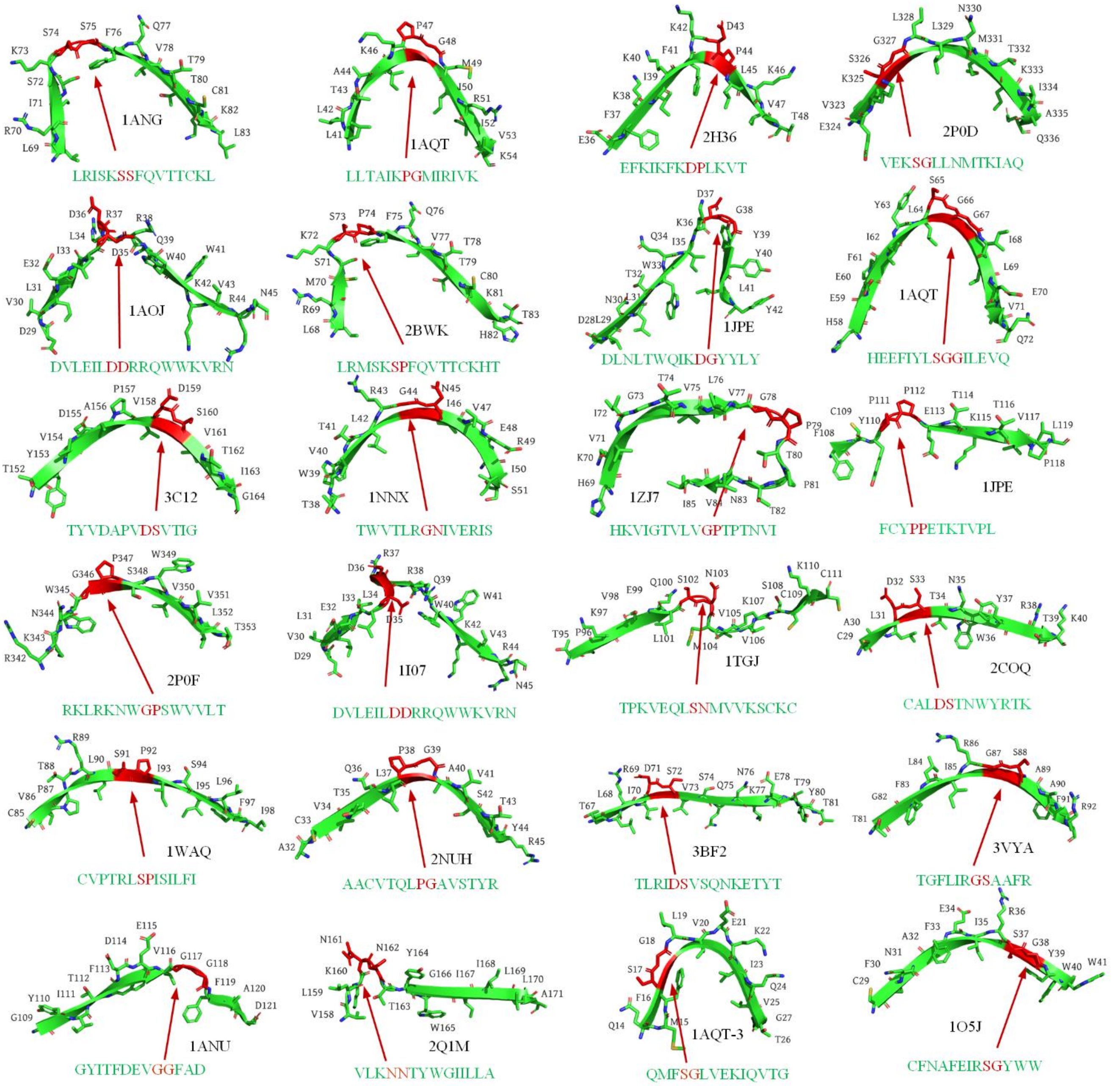

Fig.4 Long β-strands (more than 12 residues) characterized with two adjacent $R_B$ residues located at the middle of the β-strands and curved exactly at their two RB residues in the amino acid sequences.

## ACKNOWLEDGEMENTS

Lin Yang is indebted to Daniel Wagner from the Weizmann Institute of Science and Liyong Tong from the University of Sydney for their support and guidance. Lin Yang is grateful for his research experience in the Weizmann Institute of Science for inspiration. The authors acknowledge the financial support from the National Natural Science Foundation of China (Grant 21601054), Shenzhen Science and Technology Program (Grant No. KQTD2016112814303055), Science Foundation of the National Key Laboratory of Science and Technology on Advanced Composites in Special Environments, the Fundamental Research Funds for the Central Universities of China, and the University Nursing Program for Young Scholars with Creative Talents in Heilongjiang Province of China (Grants UNPYSCT-2017126).

**Additional information**

The authors declare no competing financial interests.

**Author Contributions** L.Yang, L.Ye and X.H. formulated the study. X.M., L.Yang, C.H. and L.S. conducted the MD simulation. L.Yang, X.M., C.H., L.S., L.L. and J.L. analyzed the PDB data and coded the protein folding codes. L.Yang, X.M., C.H. and L.S collected and analysed the electric charge and rotational resistance data of side-chains. C.H. wrote programs. L.Yang, L.Ye and X.H. wrote the paper, and all authors contributed to revising it. All authors discussed the results and theoretical interpretations.

**References**

1 Dill, K. A. & MacCallum, J. L. The Protein-Folding Problem, 50 Years On. *Science* **338**, 1042-1046, doi:10.1126/science.1219021 (2012).

2 Lednev, Igor K. Amyloid Fibrils: the Eighth Wonder of the World in Protein Folding and Aggregation. *Biophysical Journal* **106**, 1433-1435, doi:10.1016/j.bpj.2014.02.007 (2014).

*Corresponding author. E-mail address: linyang@hit.edu.cn (Lin Yang) [1]These authors contributed equally to this work.

3 Alberts B, J. A., Lewis J, Raff M, Roberts K, Walters P. *Molecular Biology of the Cell*. 4th edn, (Garland Science, 2002).
4 Grishin, N. V. Fold Change in Evolution of Protein Structures. *Journal of Structural Biology* **134**, 167-185, doi:https://doi.org/10.1006/jsbi.2001.4335 (2001).
5 van den Berg, B., Wain, R., Dobson, C. M. & Ellis, R. J. Macromolecular crowding perturbs protein refolding kinetics: implications for folding inside the cell. **19**, 3870-3875, doi:10.1093/emboj/19.15.3870 (2000).
6 van den Berg, B., Wain, R., Dobson, C. M. & Ellis, R. Macromolecular crowding perturbs protein refolding kinetics: implications for folding inside the cell. *The EMBO Journal* **19**, 3870-3875, doi:10.1093/emboj/19.15.3870 (2000).
7 Anfinsen, C. B. Principles that Govern the Folding of Protein Chains. *Science* **181**, 223-230, doi:10.1126/science.181.4096.223 (1973).
8 Leopold, P. E., Montal, M. & Onuchic, J. N. Protein folding funnels: a kinetic approach to the sequence-structure relationship. *Proceedings of the National Academy of Sciences* **89**, 8721-8725, doi:10.1073/pnas.89.18.8721 (1992).
9 Zwanzig, R., Szabo, A. & Bagchi, B. Levinthal's paradox. *Proceedings of the National Academy of Sciences* **89**, 20, doi:10.1073/pnas.89.1.20 (1992).
10 Dill, K. A. & Chan, H. S. From Levinthal to pathways to funnels. *Nature Structural Biology* **4**, 10, doi:10.1038/nsb0197-10 (1997).
11 Walther, K. A. *et al.* Signatures of hydrophobic collapse in extended proteins captured with force spectroscopy. *Proceedings of the National Academy of Sciences of the United States of America* **104**, 7916-7921, doi:10.1073/pnas.0702179104 %J Proceedings of the National Academy of Sciences (2007).
12 Chaplin, M. (www1.lsbu.ac.uk/water/water_structure_science.html. ).
13 Soares, C. M., Teixeira, V. H. & Baptista, A. M. Protein structure and dynamics in nonaqueous solvents: insights from molecular dynamics simulation studies. *Biophys J* **84**, 1628-1641, doi:10.1016/S0006-3495(03)74972-8 (2003).
14 Zhang, L. *et al.* Mapping hydration dynamics around a protein surface. *Proceedings of the National Academy of Sciences of the United States of America* **104**, 18461-18466, doi:10.1073/pnas.0707647104 (2007).
15 Lin, Y. *et al.* Universal Initial Thermodynamic Metastable state of Unfolded Proteins. *Progress in biochemistry and biophysics* **46**, 8, doi:10.16476/j.pibb.2019.0111 (2019).
16 Qiao, B., Jiménez-Ángeles, F., Nguyen, T. D. & Olvera de la Cruz, M. Water follows polar and nonpolar protein surface domains. *Proceedings of the National Academy of Sciences* **116**, 19274-19281, doi:10.1073/pnas.1910225116 %J Proceedings of the National Academy of Sciences (2019).
17 McPherson, A. & Gavira, J. A. Introduction to protein crystallization. *Acta Crystallogr F Struct Biol Commun* **70**, 2-20, doi:10.1107/S2053230X13033141 (2014).
18 Berman, H., Henrick, K. & Nakamura, H. Announcing the worldwide Protein Data Bank. *Nature Structural Biology* **10**, 980, doi:10.1038/nsb1203-980 (2003).
19 Berman, H., Henrick, K., Nakamura, H. & Markley, J. L. The worldwide Protein Data Bank (wwPDB): ensuring a single, uniform archive of PDB data. *Nucleic Acids Research* **35**, D301-D303, doi:10.1093/nar/gkl971 (2007).
20 Brooks, B. R. *et al.* CHARMM: the biomolecular simulation program. *J Comput Chem* **30**, 1545-1614, doi:10.1002/jcc.21287 (2009).
21 Yang, L. *et al.* Structure relaxation via long trajectories made stable. *Phys Chem Chem Phys* **19**, 24478-24484, doi:10.1039/c7cp04838f (2017).
22 Uversky, V. N. Intrinsically Disordered Proteins and Their "Mysterious" (Meta)Physics. *Frontiers in Physics* **7**, doi:10.3389/fphy.2019.00010 (2019).

23 Compiani, M. & Capriotti, E. Computational and Theoretical Methods for Protein Folding. *Biochemistry* **52**, 8601-8624, doi:10.1021/bi4001529 (2013).
24 Callaway, D. J. Solvent-induced organization: a physical model of folding myoglobin. *Proteins* **20**, 124-138, doi:10.1002/prot.340200203 (1994).
25 Rose, G. D., Fleming, P. J., Banavar, J. R. & Maritan, A. A backbone-based theory of protein folding. *Proceedings of the National Academy of Sciences of the United States of America* **103**, 16623-16633, doi:10.1073/pnas.0606843103 (2006).
26 Voet D, V. J., Pratt. *Fundamental of Biochemistry: Life at the Molecular Level* 4th edn, (Wiley & Sons, 1999).
27 Pace, C. N., Shirley, B. A., McNutt, M. & Gajiwala, K. Forces contributing to the conformational stability of proteins. **10**, 75-83, doi:10.1096/fasebj.10.1.8566551 (1996).
28 Dill, K. A., Ozkan, S. B., Shell, M. S. & Weikl, T. R. The Protein Folding Problem. *Annual review of biophysics* **37**, 289-316, doi:10.1146/annurev.biophys.37.092707.153558 (2008).
29 Voet D, V. J., Pratt *Fundamentals of Biochemistry: Life at the Molecular Level* (Wiley & Sons, 1999).
30 A, K. *et al.* Cooperativity in protein-folding kinetics. (1993).
31 Muoz, V., Thompson, P. A., Hofrichter, J. & Eaton, W. A. J. N. Folding dynamics and mechanism of β-hairpin formation. **390**, 196-199 (1997).
32 Petrovich, M., Jonsson, A. L., Ferguson, N., Daggett, V. & Fersht, A. R. phi-Analysis at the experimental limits: Mechanism of beta-hairpin formation. *Journal of Molecular Biology* **360**, 865-881, doi:10.1016/j.jmb.2006.05.050 (2006).
33 Dobson, C. M. Protein folding and misfolding. *Nature* **426**, 884-890, doi:10.1038/nature02261 (2003).
34 Clarke, D. T., Doig, A. J., Stapley, B. J. & Jones, G. R. The α-helix folds on the millisecond time scale. *Proceedings of the National Academy of Sciences of the United States of America* **96**, 7232-7237 (1999).
35 Chen, E. H. L. *et al.* Directly monitor protein rearrangement on a nanosecond-to-millisecond time-scale. *Scientific Reports* **7**, 8691, doi:10.1038/s41598-017-08385-0 (2017).
36 Dyson, H. J., Wright, P. E. & Scheraga, H. A. The role of hydrophobic interactions in initiation and propagation of protein folding. *Proceedings of the National Academy of Sciences of the United States of America* **103**, 13057-13061, doi:10.1073/pnas.0605504103 (2006).
37 Cui, D., Ou, S. & Patel, S. Protein-spanning water networks and implications for prediction of protein–protein interactions mediated through hydrophobic effects. **82**, 3312-3326, doi:10.1002/prot.24683 (2014).
38 Wang, Q. & Smith, C. Molecular biology genes to proteins, 3rd edition by B. E. Tropp. **36**, 318-319, doi:10.1002/bmb.20196 (2008).
39 Voet D, V. J., Pratt CW. *Principles of Biochemistry*. (Wiley, 2016).
40 Eisenberg, D. Three-dimensional structure of membrane and surface proteins. *Annual review of biochemistry* **53**, 595-623 (1984).
41 Marcos, E. *et al.* Principles for designing proteins with cavities formed by curved β sheets. *Science* **355**, 201-206, doi:10.1126/science.aah7389 %J Science (2017).
42 Heinig, M. & Frishman, D. STRIDE: a web server for secondary structure assignment from known atomic coordinates of proteins. *Nucleic acids research* **32**, W500-W502, doi:10.1093/nar/gkh429 (2004).

43 Eudes, R., Le Tuan, K., Delettré, J., Mornon, J.-P. & Callebaut, I. A generalized analysis of hydrophobic and loop clusters within globular protein sequences. *BMC Structural Biology* **7**, 2, doi:10.1186/1472-6807-7-2 (2007).

**Supplementary Information**

**Note S1. Protein structures samples**

| | | | | | | | | | |
|---|---|---|---|---|---|---|---|---|---|
| 134L | 1KP4 | 1Z3L | 3C3G | 4IC9 | 1FDD | 1R75 | 2OAI | 3OZZ | 5HBP |
| 155C | 1KSM | 1Z3M | 3C4S | 4IDL | 1FE3 | 1R9H | 2OJR | 3P2X | 5HBQ |
| 1A0B | 1KTH | 1Z3P | 3C5K | 4IP1 | 1FE5 | 1RAQ | 2OLI | 3PAZ | 5HDG |
| 1A18 | 1KXI | 1ZIA | 3C7I | 4IP6 | 1FER | 1RBI | 2ON8 | 3Q1D | 5HJC |
| 1A6F | 1KXW | 1ZIB | 3C97 | 4IPF | 1FES | 1RBW | 2OPY | 3Q4Y | 5HKN |
| 1AA2 | 1KXX | 1ZJ7 | 3CE8 | 4JHB | 1FEV | 1RBX | 2OQK | 3Q7Y | 5HMB |
| 1AB0 | 1KXY | 1ZPA | 3CEC | 4JJD | 1FF2 | 1RDS | 2OSN | 3QD7 | 5HPA |
| 1ACD | 1L5D | 208L | 3CQ1 | 4JZ3 | 1FIK | 1REX | 2OUB | 3R3K | 5HQI |
| 1ACF | 1L7L | 2A3G | 3CR2 | 4JZP | 1FKD | 1RFP | 2OUM | 3R5P | 5HVZ |
| 1AE2 | 1LAC | 2A7B | 3CR5 | 4K1V | 1FKF | 1RH2 | 2OXK | 3RH1 | 5I72 |
| 1AF5 | 1LCJ | 2A9I | 3CTG | 4K59 | 1FKK | 1RLK | 2OYZ | 3RHE | 5J6X |
| 1AH0 | 1LEA | 2A9Q | 3CYI | 4KGT | 1FKV | 1RMD | 2OZ9 | 3RNT | 5JOE |
| 1ALB | 1LGP | 2AIF | 3DJN | 4KUO | 1FLQ | 1RNN | 2P0D | 3RSD | 5K2L |
| 1ANG | 1LKL | 2APB | 3DJU | 4KV4 | 1FLU | 1RNQ | 2P0F | 3RSK | 5KAZ |
| 1ANU | 1LOZ | 2AZ8 | 3DML | 4LBB | 1FLW | 1RNU | 2P1X | 3RTO | 5KM6 |
| 1AOJ | 1LPI | 2AZ9 | 3DP5 | 4LFQ | 1FLY | 1RNV | 2P2E | 3RVC | 5KUE |
| 1AQT | 1LRA | 2AZB | 3EAZ | 4LFS | 1FN5 | 1RNW | 2P61 | 3S3Y | 5KXF |
| 1AWJ | 1LSL | 2B1Y | 3ENU | 4LJN | 1FNJ | 1RNX | 2P64 | 3S8S | 5L8Z |
| 1AYC | 1LVE | 2B29 | 3ERS | 4LTT | 1FNK | 1RNZ | 2P6V | 3S9K | 5LAW |
| 1B0T | 1LXI | 2B4A | 3ETW | 4LYO | 1FOW | 1RRI | 2P8V | 3SGP | 5LAZ |
| 1B1E | 1LYO | 2B8G | 3EZM | 4MDQ | 1FOY | 1RRY | 2PAL | 3STM | 5LN2 |
| 1B1I | 1LZ4 | 2B9D | 3F3Q | 4MJJ | 1G2S | 1RS2 | 2PK7 | 3SUL | 5M9A |
| 1B1J | 1M1S | 2BEZ | 3F45 | 4ML2 | 1GBQ | 1RSI | 2PKT | 3T1X | 5MXY |
| 1B1U | 1M4A | 2BFH | 3F8C | 4MZ2 | 1GD6 | 1RTU | 2PNE | 3T3J | 5NGN |
| 1B2O | 1M4B | 2BHK | 3FAJ | 4N0Z | 1GDC | 1RWJ | 2PPI | 3T8R | 5NWX |
| 1B6E | 1M4M | 2BHO | 3FFY | 4N5T | 1GHJ | 1RZY | 2PPN | 3T8T | 5O3A |
| 1BAS | 1MB3 | 2BO1 | 3FRV | 4N6J | 1GKH | 1S3P | 2PVT | 3UAF | 5OC4 |
| 1BEA | 1MG6 | 2BPP | 3FWU | 4NEJ | 1GMG | 1S7I | 2PW5 | 3UB2 | 5OC8 |
| 1BEL | 1MH7 | 2BQQ | 3FYR | 4NXR | 1GOD | 1SDZ | 2PWS | 3UB3 | 5OMD |
| 1BFE | 1MH8 | 2BRF | 3FZ9 | 4OOH | 1GP3 | 1SF6 | 2PYK | 3UB4 | 5PAL |
| 1BGI | 1MK0 | 2BS5 | 3FZA | 4OV1 | 1GS3 | 1SF7 | 2PZW | 3UMD | 5PAZ |
| 1BHF | 1MKU | 2BTI | 3G7C | 4OXW | 1GSW | 1SFB | 2Q1M | 3UME | 5RHN |
| 1BKF | 1ML8 | 2BWK | 3GBQ | 4OZL | 1GV2 | 1SFG | 2QAS | 3UNN | 5RNT |
| 1BKV | 1MLI | 2BWL | 3GK2 | 4P15 | 1GVP | 1SKZ | 2QDB | 3V19 | 5TAB |
| 1BM2 | 1N9N | 2BZY | 3GK4 | 4P2P | 1GXT | 1SNP | 2QHE | 3V1G | 5U9U |
| 1BMG | 1N9O | 2CDS | 3GKY | 4P7U | 1H2P | 1SNQ | 2QHW | 3VXW | 5UEP |
| 1BOO | 1NEH | 2COQ | 3GLW | 4P9E | 1HDO | 1SSC | 2QIU | 3VYA | 5UER |
| 1BPQ | 1NEQ | 2CW4 | 3GM2 | 4P9V | 1HE7 | 1SV9 | 2QNW | 3WIT | 5UES |
| 1BQK | 1NKO | 2CXY | 3GM3 | 4PAZ | 1HEH | 1T00 | 2QR3 | 3WRP | 5UET |
| 1BQR | 1NLO | 2D4L | 3GQU | 4PE7 | 1HEY | 1T1D | 2QVT | 3WUN | 5UEV |
| 1BU3 | 1NN7 | 2D4M | 3GSP | 4PTA | 1HFX | 1T2J | 2R2Y | 3WVT | 5UEY |
| 1BXY | 1NNX | 2D58 | 3H33 | 4QYW | 1HIK | 1TAY | 2R34 | 3WW5 | 5USV |
| 1BYL | 1NXT | 2DB7 | 3H9W | 4RLM | 1HKF | 1TCY | 2R36 | 3WX4 | 5UVS |
| 1C0B | 1NXV | 2DH1 | 3HAF | 4RLN | 1HME | 1TDY | 2R39 | 3ZEK | 5UVY |
| 1C0C | 1O2E | 2DP9 | 3HAK | 4RUD | 1HQ8 | 1TEN | 2REA | 3ZK0 | 5UVZ |
| 1C76 | 1O3X | 2DRT | 3HEY | 4TQN | 1HQB | 1TGJ | 2REY | 4AHG | 5VEA |
| 1C9H | 1O41 | 2DU9 | 3HFF | 4TS8 | 1HRC | 1TNS | 2RKN | 4AHI | 5VR3 |
| 1C9V | 1O42 | 2DYQ | 3HFN | 4UNG | 1HRQ | 1TS9 | 2RLN | 4AHN | 5WOR |
| 1CDP | 1O43 | 2EFF | 3HJX | 4UNV | 1HSQ | 1TUW | 2RNS | 4AQI | 5WPB |
| 1CDT | 1O44 | 2EG2 | 3HNU | 4UTQ | 1I07 | 1TVQ | 2SAK | 4AQJ | 5WRB |
| 1CHN | 1O45 | 2EH9 | 3HQA | 4WBF | 1I0C | 1TXA | 2UUX | 4AYA | 5XUK |
| 1CI6 | 1O47 | 2EQL | 3HQB | 4WO6 | 1I0V | 1U36 | 2UZR | 4B9J | 5XWE |
| 1CKA | 1O4A | 2ES9 | 3I3Z | 4X1L | 1I3Z | 1U3J | 2V5F | 4BFN | 5YDN |
| 1CM3 | 1O4C | 2EYW | 3I7T | 4X1M | 1I9E | 1U3Y | 2V6H | 4BS7 | 5YI1 |
| 1COF | 1O4E | 2F3L | 3I7W | 4XC4 | 1IDI | 1U3Z | 2VJW | 4CGQ | 5YM7 |
| 1CS3 | 1O4G | 2F4G | 3I7X | 4XO2 | 1IGU | 1U42 | 2VKS | 4CPV | 5ZND |
| 1CU0 | 1O4H | 2F4K | 3I7Y | 4XXL | 1IIZ | 1U68 | 2VP7 | 4CQI | 6B25 |
| 1CYI | 1O4I | 2FB0 | 3IE9 | 4Y92 | 1IJ0 | 1U9R | 2VSD | 4DH0 | 6BBK |
| 1D1M | 1O4J | 2FCD | 3IJU | 4YM8 | 1IJ1 | 1UHD | 2VSL | 4DH4 | 6BSY |
| 1DAQ | 1O4K | 2FD2 | 3IN2 | 4YOP | 1IJ2 | 1UIA | 2W1R | 4DP0 | 6CE6 |
| 1DCF | 1O4L | 2FKE | 3INC | 4YX5 | 1IKL | 1UIC | 2W51 | 4DP1 | 6CE8 |
| 1DDV | 1O4M | 2FKL | 3IRC | 4ZC3 | 1IOR | 1UID | 2WOR | 4DP2 | 6CEA |
| 1DDW | 1O4N | 2FLS | 3J4G | 5AEF | 1IOS | 1UIE | 2WOS | 4DP4 | 6CEC |
| 1DKJ | 1O4O | 2FLT | 3JTE | 5AFG | 1IOT | 1UIF | 2WQ0 | 4E1P | 6CED |
| 1DMM | 1O4Q | 2FMB | 3JZR | 5AI2 | 1IR7 | 1UIG | 2WQ3 | 4E1R | 6CEE |
| 1DMN | 1O4R | 2FO3 | 3K0X | 5B1F | 1IR8 | 1UKU | 2WQH | 4EC2 | 6CEF |
| 1DMQ | 1O5J | 2FQL | 3K63 | 5B1G | 1IR9 | 1UPJ | 2X44 | 4EES | 6DXH |
| 1DPY | 1O7Z | 2FTB | 3K6D | 5B3Q | 1IRQ | 1UPR | 2X9C | 4ET8 | 6DXR |
| 1DQ7 | 1OAP | 2GSP | 3K9I | 5B52 | 1IRW | 1UVY | 2XBD | 4ET9 | 6EKB |
| 1DYZ | 1OB6 | 2GV2 | 3KD0 | 5B79 | 1ISU | 1UW3 | 2XCZ | 4ETA | 6EVL |
| 1DZ0 | 1OD7 | 2H2B | 3KHQ | 5BMH | 1IYU | 1UW7 | 2XDY | 4ETB | 6EVM |
| 1E5B | 1ODA | 2H36 | 3KLU | 5BP0 | 1IZA | 1UWM | 2XFE | 4ETD | 6FCE |
| 1E6K | 1OOI | 2H46 | 3KMJ | 5C39 | 1J2V | 1V07 | 2XKH | 4ETE | 6FGL |
| 1E6L | 1OPY | 2H8A | 3KQ6 | 5C4P | 1J4H | 1V46 | 2XMU | 4EWW | 6FGT |
| 1E6M | 1OR5 | 2H8V | 3KQI | 5C68 | 1J4I | 1VAT | 2YHN | 4EWZ | 6FGU |
| 1E97 | 1P0R | 2H95 | 3KTP | 5C6X | 1J73 | 1VED | 2YWZ | 4EX0 | 6FH6 |
| 1EA2 | 1P65 | 2HB4 | 3KU7 | 5CB9 | 1J7Z | 1VER | 2YXY | 4F1A | 6FH7 |
| 1ECW | 1P9G | 2HC8 | 3KVT | 5CO9 | 1J81 | 1VHF | 2Z44 | 4F2E | 6FXF |
| 1ED1 | 1PAL | 2HPL | 3L1M | 5CPV | 1J82 | 1VIH | 2Z7J | 4F68 | 6GF0 |
| 1EFQ | 1PAZ | 2HWV | 3L6W | 5CUL | 1JC7 | 1VMG | 2ZGD | 4F69 | 6H0K |
| 1EIF | 1PBI | 2I9V | 3LHC | 5D53 | 1JDL | 1W2L | 2ZHH | 4F6D | 6H0L |
| 1EIG | 1PI8 | 2IGP | 3LJM | 5D54 | 1JER | 1W6X | 3A0D | 4FC1 | 6HA4 |
| 1EMN | 1PKS | 2IIY | 3LLH | 5DKN | 1JHC | 1WAQ | 3AOS | 4FDX | 6I3S |
| 1EN2 | 1P08 | 2INT | 3LMF | 5DM9 | 1JKD | 1WHI | 3AOV | 4FE6 | 6I5A |
| 1ENV | 1PZ4 | 2J5A | 3LRO | 5DZD | 1JOI | 1WJG | 3A7L | 4G3O | 6IFH |
| 1EOQ | 1PZA | 2JIN | 3LVE | 5EOZ | 1JON | 1WJX | 3ADY | 4G4W | 6INS |
| 1EOT | 1PZB | 2JK7 | 3LYO | 5E4P | 1JPE | 1WRP | 3AZ5 | 4G4X | 6IQC |
| 1EOW | 1PZC | 2JP8 | 3LZ2 | 5E4X | 1JPO | 1WY9 | 3B7X | 4GBC | 6JQ4 |
| 1EQV | 1Q4R | 2JVH | 3M4T | 5EE2 | 1JZA | 1XW3 | 3B84 | 4GBN | 6MHN |
| 1EV3 | 1Q4V | 2KRI | 3MAZ | 5EH4 | 1K1Z | 1XW4 | 3BF2 | 4GFY | 6MPM |
| 1F1W | 1Q7O | 2LYO | 3MCD | 5EN9 | 1K40 | 1Y49 | 3BI9 | 4GQV | 6NF3 |
| 1FAO | 1Q8B | 2M6D | 3MF8 | 5ER4 | 1K58 | 1Y9T | 3BIA | 4GSP | 6PAZ |
| 1FAV | 1QFI | 2M6E | 3MYA | 5ET3 | 1K5A | 1YAT | 3BJO | 4HB6 | 6Q5K |
| 1FAZ | 1QKF | 2M6G | 3NGR | 5FD1 | 1KC2 | 1YEA | 3BKS | 4HE7 | 6QQJ |
| 1FB7 | 1QL2 | 2M6H | 3NRW | 5FG4 | 1KCQ | 1YEB | 3BOV | 4HMB | 6RLX |
| 1FB8 | 1QLS | 2MSS | 3NZ3 | 5G25 | 1KHO | 1YGT | 3BY5 | 4HP9 | 6RNT |
| 1FD2 | 1QPV | 2NUH | 3O7O | 5GSP | 1KH8 | 1YMV | 3BYR | 4HRS | 7PAZ |
| 1FD8 | 1QSU | 2OOP | 3O8W | 5HOY | 1KJT | 1YV7 | 3BZS | 4HV2 | 8PAZ |
| 1FDA | 1QTO | 2OOQ | 3ONH | 5HBL | 1KM8 | 1YVS | 3C12 | 4I3I | 9RAT |
| 1FDB | 1R26 | 2O1N | 3OSE | 5HBO | 1KM9 | 1Z21 | 3C1R | 4IAS | 9RNT |

Table. S1. Randomly selected 1000 small protein structures in PDB

**Note S2. Amino acid sequences of β-turns of the 1000 protein samples**

PDBID:1AA2

AGYPNV NFT RDG RPDLI KKSN TKL VDHP

PDBID:1ACF

NL GAVT LDG SAGF AG DDR GS

TSK NEKI

PDBID:1AOJ

NSS MKD ASG NNI TPE NSSE MKD

DDRR NASG

PDBID:1AYC

RRW HPNI VDG KSNPG NG TGDY LYGG

HHGQ KNG

PDBID:1BU3

AFSGILA AADS DQDK SAGA DSDG

PDBID:1CKA

DEE KKG KPEEQ DSEG

PDBID:1FB7

LWQR GG DTGA IG CG PTPVN

PDBID:1FB8

SLGT GGLVK RN KDQMS YSQERV PF

PDBID:1FES

CSGCSGA

PDBID:1NXV

PD LMLPEID AD KPF

PDBID:1O42

EEW KI NAENPRG SETTKGA NAK LDSG SR

ADGLCHR

PDBID:1O4C

EEW KI NAENPRG SETTKGA NAK LDSG SR

ADGLCHR

PDBID:1O4J

EEW KI NAENPRG SETTKG NAK LDSG SR

ADGLCHR

PDBID:1O4N

SIQAEEW KI NAENPRG SETTKGA NAK LDSG SR

ADGLCHR

PDBID:1O4O

SIQAEEW KI NAENPRG SETTKGA NAK LDSG SR

ADGLCHR

PDBID:1OD7

PK LSG PADV VSNG DYSGKR SGD DKRFRGR

KD DS DVKG

PDBID:1ODA

PDBID:1FNK
RDT TPDL LSGWQYV VTGGLKK
PDBID:1FOY
TFIT PDL SMG
PDBID:1GBQ
ADD KRGD DQN NG KNY
PDBID:1GVP
KPSQA SRQG NEYP DEGQ GQFG DRL
PDBID:1J2V
EG YDVP NED
PDBID:1J4I
DGRT KRG EDG GKQEVI SVG ATG PGI
PPH
PDBID:1J82
NLTKDR CKNG TGSS YP GNPY
PDBID:1K5A
DAKPQGR LTSPCKD ENKN REN WPPC NG QSA
PDBID:1LEA
FRS VSGAS
PDBID:1LGP
RLGAEEGE KR RRGCD DEKSG STSG VKKQ QTG
YRKNE
PDBID:1MH8
VPAR GCY QG KGRN NI
PDBID:1MKO
TLNN KD HSS GNV PYE NSKIN
PDBID:1NLO
LYD TET KKG NTEGD LTTG
PDBID:1NNX
RDD DD AS WNGV TPKD WN
PDBID:1O4E
EEW KI NAENPRG SETTKGA NAK LDSG SR
ADGLCHR
PDBID:1OPY
DPFGQ SHNG NG DEHG
PDBID:1P9G
CPRP NAG IYG GAGN
PDBID:1RDS
GS DD DYEG MSDY GDD HTGASGD D
PDBID:1S3P
AADS GLKKK DKDK SSDA DKDG
PDBID:1SKZ
PADV DYSGKR SGDPKLV RGR MDHK KD DS
DVKG
PDBID:1PKS
LYD REE HLG SDG
PDBID:1PO8
GCY IDGC EP DSKD
PDBID:1PZA
AE PA NPG VDK IKDM PEGA KINE
CTPH GDSP
PDBID:1PZB
AE PA NPG VDK IKDM PEGA KINE
CTPH GDSP
PDBID:1RNQ
DSSTS NLTKDR CKNG TGSS YP GNPY
PDBID:1RNV
NLTKDR CKNG TGSS YP GNPY
PDBID:1RNX
DSSTS NLTKDR CKNG TGSS YP GNPY
PDBID:1RRI
YNR ENPPI
PDBID:1UIG
LDN NF TQA NTDG GILQ SRWW DGRTPGS
NLCN SSD DGN KGTD RGC
PDBID:1UW7
PNN TTQTAC NSKGG HQDL KSDG TPKG KGL
PDBID:1WJG
FK PDYL GTRG SLFL
PDBID:2BEZ
NVLY ISGI NESL
PDBID:2BPP
IPSS NN GCY KVL NPYTN NN SSEN
NL
PDBID:2BTI
VG GDE GN PKEV VG GDE GN
PKEV
PDBID:2CXY
GK PEV
PDBID:2DH1
QGDV LSNG NLNKVNS SSATG IKDY PEDT PAGS
GGGTLGV DTKG NSEYN VSGA DSSTK NDNG SLGG
QGDV LSNG NLNKVNS SSATG IKDY PEDT PAGS
GGGTLGV DTKG NSEYN VSGA DSSTK NDNG SLGG

PEGS IITD CPH SRYG LEP PEG RLTN
IDI CPN DKLG
PDBID:1T1D
SG FPDT PLRN GGR PVNV
PDBID:1TGJ
WKW PK PYLRSA NPEAS GR
PDBID:1TS9
WIG SPN VG TQN TEKG KRGR KG
PDBID:1UPJ
LWQRP GG DTGA GIGG CG TPVN QIG
PDBID:1UVY
NGI GPNA ANM
PDBID:1VIH
HK GKSG SEKS
PDBID:2AIF
ADAEP KDGS
PDBID:2AZ9
LWKRP GG DTGA GIGG CG GPT QIG
PDBID:2BHK
WDDW PL EFPL DPEST DSAN
PDBID:2BHO
NEND DGH HQQ LGAD FGEHWPA DG LVGL
PDBID:2BO1
ARNT AGES
PDBID:2EFF
GG ENG KQG GD DRSG
PDBID:2FCD
DKKG KD DSSLRDAS
PDBID:2H46
HPW KI HDG ESAPG GN DGAG RST
SRNQ
PDBID:2HC8
DG AVG RPG KG FGA
PDBID:2HPL
NPSD IGN LPVRG GET PKKA
PDBID:2JVH
AMVSI KDS PMDV
PDBID:2M6H
GCPDP
PDBID:2QNW
NPES
PDBID:2VSD
PSQG SLG PRM NG DKEQ SEPL

SEGNPAI QGDV LSNG NLNKVNS SSATG PEDT PAGS
GGGTLGV DTKG NSEYN VSGA DSSTK NDNG SLGG
SEGNPAI QGDV LSNG NLNKVNS SSAT IKDY PEDT
PAGS GGGTLGV DTKG NSEYN VSGA DSSTK NDNGD
DLKAKL SLGG
PDBID:2FD2
IKCK EVCPVD PN HPDEC EDEV DGVKGK
PDBID:2FLT
QDRLT PAGN GG VEYG
PDBID:2LYO
LDN NF TQA NTDG GILQ SRWW DGRTPGS
NLCN SSD DGN KGTD IRGC
PDBID:2O1N
GCY GWG NG LY
PDBID:2OUM
KELA LPHG DKTG ASF AHKPEGA TF TMG
PDBID:2POD
GG GN LRGA GRHLS RRN TIPG SDH
PDBID:2PPN
DGRT KRG EDG GKQEVI SVG ATG PGI
PPH
PDBID:2QR3
NHFS MNFTSNE YRDL PW
PDBID:2UZR
GKYIK NDG KER VAQ ERPRPN WT
PDBID:2V5F
DPEH HHHH
PDBID:2W1R
RD GS RS SDA NG NG KDQ
PDBID:3F45
TAAD DKDK SSDA DKDG
PDBID:3FZ9
NP KTWC TVP GG
PDBID:3HFF
DGP KESNG HEEV DKDG SVI GDH IIG
DLGK
PDBID:3HFN
VTG DPT DRQ KQA SLP VTGD DPT
DRQ
PDBID:3KDO
AGDK ATWC SN VDDS SMP KG
PDBID:3LYO
LDN NF TQA NTDG GILQ SRWW DGRTPGS

PDBID:2YWZ
PR ETGE LGS GR KGSK SDL SPSCYSYP
PDBID:2ZGD
DVAA DKNG DKFG
PDBID:3BJO
LDYI EE FKGKYE PQRG QS
PDBID:3C1R
NE KTYC KLK TVP NG PIL
PDBID:3CE8
ND LEFL RE KHN
PDBID:3DJU
NHK PSE PY GEDG
PDBID:3F3Q
HHH ATWC YPQA VDEL AMP NG
PDBID:3GM2
RTDD VM AVT
PDBID:3H33
TRIG IDGF GKGC
PDBID:3HAF
VGGLGGY
PDBID:3K6D
NQ VSDRIPGS NENSG DREAIP DENG
PDBID:3KQI
GSMA CVCR DVTR CDACK KKKR
PDBID:3KTP
HPT NWPP PG WKGL
PDBID:3MCD
ERTL LPYM ERTL LPY
PDBID:3NRW
TDA ARE DEDL
PDBID:3T8R
KEGQ RGAKG RVP IESE
PDBID:4B9J
KEG PENV VY TADDSQ DDKG KD TPN
PDBID:4CPV
FAGVLN AADS GLTSK DQDK FKADA DSDG
PDBID:4E1P
DDFDGSG DG DDFDG DG
PDBID:4F2E
NG MGG PE KKSV PDFG PMG GMNMM
PDBID:4G3O

NLCN SSD DGNGMNAW KGTD RGC
PDBID:3O8W
SDD KFGD IRTG
PDBID:3Q4Y
VPSR GCY KG QG KGDN NI
PDBID:3RVC
MLI GP QAI FD EEG LPSL
PDBID:3UB4
SRWSKD GAI QALS PGAEG LGL GRG
PDBID:4AHN
DNS DAKPQGR TSPCKD ENKN RENL WPPC NG
QSIFH
PDBID:4BFN
PSSAS DGS NIAY DGSD SGS VKGI SG
PDBID:4CQI
KEGSTV
PDBID:4DHO
DGRT KRG EDG GK SVG ATG PGI
PPH
PDBID:4F69
FGFKGV
PDBID:4G4X
DYDS NANS KEAPVNP
PDBID:4K1V
DPFGQ TGP SHNG NG DEHG
PDBID:4LBB
IPA IAGE QG IPA IAGE QG
PDBID:4MJJ
GL TLNP GI EDE
PDBID:4UNV
SLGQ TSSDVGGY HAGK EVN PSGVPDR GN SGL
SD
PDBID:4X1M
YKDSP VPGK DRSS GG LLQEF SAPT TDK
GV
PDBID:4Y92
TTTD KKG NTEG SLTTG
PDBID:5B1F
LDN NF TQA NTDG GILQ SRW DGRTPGS
LCN SSD DGNGMNAW KGT RGC
PDBID:5B3Q

FPQV
PDBID:4G4W
DYDS NANS KEAPVNP
PDBID:4GBC
TSI TSI
PDBID:4I3I
AFGS GDG VIG SPEF DYQM SGD
PDBID:4NXR
TAAD DG KETG KAG NN
PDBID:4OOH
DSST NLTKDR CKNG TGSS YP GNPY
PDBID:4X1L
ADG YKDSP VPGK GG LLQDGEF KPT TDK
GV
PDBID:4ZC3
DMN
PDBID:5AI2
KICG EDA SPGT PDDW CPICG
PDBID:5C68
PD PG AEPP GAGE GPGG
PDBID:5D54
TSI FVNQ
PDBID:5DZD
PEG TVDG HNRR PRTG PEG TVDG HNRR
PRTG
PDBID:5E4X
GDDGKT MSDA PQDNV
PDBID:5HOY
SP AVRG AIN AADD SESG TEEEFV PFYE
NDSG PY
PDBID:5I72
PQF KSCW NKG CNNH CPICK PQF KSCWF
NKG CNNH CPICK
PDBID:5JOE
LKD YPEI GT SDK GD KN GP
PDBID:5L8Z
TDVL AGF NPSTG
PDBID:5RHN
RPGG KEIP DDQ DISP AP KKH LKKG
QSVY NWPP
PDBID:5RNT
CD GS GSNS NYEGFDF LSSG ENN HTGASGNN
DNGLGN GFN YLTY RPER WKD PEEDP VTDLW
GH
PDBID:5EE2
NG FNSN SDY GP GG EV DGD
LDD
PDBID:5HBL
IPE HE QAG APA ED
PDBID:5HBQ
IPE HE DQAG APA ED
PDBID:5HJC
YKP IIKHP PPDH
PDBID:5HPA
IPE HE QAG APA ED
PDBID:5HVZ
PER PER
PDBID:5KM6
KEIP DDR DISP AP LNKG QSVY HWPP
PDBID:5KUE
NG PS KPISD DRGI SPDG ND
PDBID:5KXF
GG PEVV LN LDDG QER
PDBID:5M9A
DAKPQGR GLTSPCKD ENKN REN GGSPWPPC NG
PDBID:5YDN
TADG ESQP
PDBID:6CE8
CPH PAAG DVTQ CGDCG IQE LSCY GRYING
YIDL YYCQ
PDBID:6CEA
PWCPH PAAG DVTQ CGDCG IQE LSCY GRYING
YIDL YYCQ
PDBID:6FGU
SMSV HEDAWPFL NLKLVPG IKKP EDDS
PDBID:6GFO
LDN NF TQA NTDG GILQ SRW DGRTPGS
LCN SSD NGMNAW KGTD RGC
PDBID:6HA4
KSKN NDAG KDN TYNN
PDBID:7PAZ
AE PA NPG VDK IKDM PEGA KINE
CTIH GDSP
PDBID:1ANU

PDBID:5UEY

AKK DVEALG IKHP PPDH

PDBID:5USV

TSI FVNQ

PDBID:5YM7

EDKSPDS GK

PDBID:6B25

PQENE RPG NED QD ANF QQNE VRT

KEN DG GK

PDBID:6IQC

KIP

PDBID:9RAT

DSST NLTKDR CKNG YP GNPY

PDBID:9RNT

CD GS GSNS NYEGFDF LSSG ENN HTGASGNN

PDBID:1AE2

KPSQA SRQG NEYP DEGQ GQFG DRL

PDBID:1AHO

DDVN PY PDHV

PDBID:1B1U

TPSG RLLQ PG LVTEVECN

PDBID:1BHF

EPW KNL APGNTHG SESTAGS DQNQG LDNG PR

SDGLCTR

PDBID:1BQR

KD PA APG TDK IKGM PDGA KINE

CTPH GDAP

PDBID:1CM3

TPNGLD NG AKS TQG GED

PDBID:1ECW

SVLS LRPGG TGTA

PDBID:1EIG

VSKR PENR RSTC KKG DPKQ KK

PDBID:1FD8

VM CSG PD LEKQ

PDBID:1FDB

IKCK CPVD PN HPDEC CPAQ EVW DGVKGK

PDBID:1FKK

PGDGRT KRG EDG SRDRN GKQEVI SVG ATG

PGI PPN

SVG GVPSKG DPNV GDII DPN NDL

PDBID:1AWJ

EETL DPQE DKNGHE SSYL

PDBID:1B1E

QDN DAKPQGR TSPCQD ENKN REN GGS WPPC

NG

PDBID:1BEA

GPDA PG

PDBID:1BMG

RHP DG PP NG KS KDW NSKD

VTLEQP

PDBID:1CHN

WNMPNMD DGAMSAL EA KEN PF

PDBID:1COF

KKYK NDAK PEND NGNE PDTA LNG DFSEVS

RG GA

PDBID:1DMN

DPFGQ SHNG NG DEHG

PDBID:1E6K

MPNMD DGAMSAL EAK PF

PDBID:1FD2

IKCK EVAPVD PN HPDEC CPAQ DGVKG

PDBID:1FN5

LDN NF TQA NTDA GILQ SRWW DGRTPGS

NLCN SSD DGNGMNAW KGTD RGC

PDBID:1HDO

SWDT PITG NG

PDBID:1HIK

HKCD TLCTEL DIFAASKN DTRCL PVK

PDBID:1HRQ

PAN AADD QDATKS QF ADAAGA DE QL

VNG CASW

PDBID:1IR9

LDN NF TQA NTDG GILQ SRWW DGRTPGS

NLCN SSD DGN KGTD RGC

PDBID:1JER

VG PSSP VG PANA FVNSDND VGTH

PDBID:1KCQ

RRV NNGD GN GSNS RSGR EGTE

PDBID:1L7L

PDBID:1FNJ

RDT TPDL LSGWQYV VTGGLKK PQDQI

PDBID:1GDC

CLVCS YG CAGRN NLEA

PDBID:1IIZ

AR TDK NKNG GLFQ DKYW GST GKDCN

TDD KFDAW NHS

PDBID:1IKL

CQCIKT HPKF GPHCA SDG DPKE

PDBID:1IRQ

PDK

PDBID:1IYU

KTG EVEQ SAKA SPKA KLGD EG PAAGA

PDBID:1J81

NLTKDR CKNG TGSS YP GNPY

PDBID:1K40

PAS

PDBID:1LKL

EPW KNL APGNTHG SESTAGS QNQ LDNG PR

SDGLCTR

PDBID:1LRA

CD GS GSNS NYEG LSSG ENN HTGASGNN

PDBID:1N9O

ATLPDC GPDEVLG GEGT RKDG TPDG

PDBID:1NEQ

NEKARD LER IWPSRYQ

PDBID:1NKO

EGM DSD NPAW RDR RDA GN

PDBID:1O4M

EEW KI NAENPRG SETTKGA NAK LDSG SR

ADGLCHR

PDBID:1O7Z

CIS NPRS SQFCP KKKGE NPES CISI SQFCP

KKGE NPES

PDBID:1P65

SDSG SDSG

PDBID:1PZ4

GIRM PA GG LKNV AE GALP

PDBID:1QLS

RDGNNT TE DLDS

PDBID:1R9H

TPKK GG TTG ENG GRGNVI DA PPK

ANNEAG NPGD YGP REH NS VNT PNNV

VPGTYGNN

PDBID:1LZ4

MDG GY TRA AGDR GIFQ SRY DGKTPGA

AAH QDN DPQ QNR VQGC

PDBID:1M4B

NYKNPK KKA

PDBID:1M4M

QIW FKNW CACT TENEPDL FFCF EPDD SPGC

VKK

PDBID:1ML8

MQA GL ASG GNSGDKA QD RDL

KAV

PDBID:1N9N

ATLPDC GEGT RKDG TPDG

PDBID:1OOI

FNF NKKG ADGQ

PDBID:1P0R

RLG NTDD QTG RWNKI KWYT KDHV LGDYE

HDG

PDBID:1PAZ

AE PA NPG VDK IKDM PEGA KINE

CTPH GDSP

PDBID:1PBI

CDT SNPP SNPP QCHN KNKSA CDT SNPP

SNPP QCHN KN

PDBID:1QPV

KKYK NDAK NGNE PDTA LNGV DFSEVS

PDBID:1RNZ

DSSTS NLTKDR CKNG TGSS YP GNPY

PDBID:1RRY

NVIDT YNR ENP

PDBID:1RSI

NVIDT YNR ENPPI

PDBID:1SFB

LDN RG NF TQA NTDG YGILQ NSRWWC

DGRTPGS NLCN LLSSDITA VSDG CKGT WIR

PDBID:1U3Z

DRT VTG KD EENG RQF YKDVNI RKSDL

PDBID:1U42

DRT CVTG QKDDI EGV SPTDV YKDVNI KSDL

PDBID:1UHD

GG
PDBID:1RBX
DSST NLTKDR CKNGQTN TGSS YP GNPY
PDBID:1SF6
GRCELAAA LDN RG NF TQA NTDG GILQ
SRWWC DGRTPGS NLCN KGT WIRGC
PDBID:1SF7
LA LDN RG NF TQA NTDG GILQ
SRWW DGRTPGS NLCN SSDITAS
PDBID:1TNS
LPGLP RAGVKG GEI TSLGYF
PDBID:1U36
DRT VTG KD RQF YKDVNI RKSDL
PDBID:1VER
PR ETGE DA LGS GR KGSK RDL
RLS
PDBID:1W6X
SKL KAG KD RG SKL KAG KD
RG
PDBID:1WRP
SPYS
PDBID:1XW3
GAQG PDL
PDBID:1XW4
GAQG PDL
PDBID:1YAT
PGDGAT KTG ENG GV SVG PRG PGL
PPN
PDBID:1ZIB
KD PA APG TDK IKGM PDGA KINE
CTPH GDAP
PDBID:1ZPA
LWKRP GG DTGA IG CG GPTPVN
PDBID:2A3G
ASV
PDBID:2BS5
GTVP NG DGK GD GS GT DGNG
PDBID:2D58
SSD NGNG
PDBID:2DB7
SGGYFD DASD DASD
PDBID:2DU9
GTLS RGI

REG VNNG KRGE AIG DEKN LNY
PDBID:1UKU
RLIA EG YDVP
PDBID:1UPR
DPNL DSSGLR GH DSRE LPSY APRGRRF PG
PDBID:1V07
VN FGFKGV
PDBID:1V46
NAFT
PDBID:1YV7
TRDVD HCK KGI TSRPCK NQ
PDBID:2B4A
CD VD QTKQ SSEH KPF
PDBID:2B9D
CAYCE CAYCE
PDBID:2CDS
LDN NF TQA NTDG GILQ SRW DGRTPGS
LCN SSD DGN KGTD IRGC
PDBID:2DP9
GG KDE RRPGR LSEV
PDBID:2EG2
FDF RVGD IRTG
PDBID:2EQL
EMDG NF TRA NANG GLFQ NKWW KRSS
NACN DPK KDK
PDBID:2ES9
IKN STY
PDBID:2GSP
CD GS GSNS NYEG LSSG ENN HTGASGNN
PDBID:2H8V
VNE SL DDEA LPT
PDBID:2INT
TLCTEL DIFAASKN DTRCL
PDBID:2KRI
VKKA QG KNG LHG KEKK DG PKCF
SSLAF DASDV GPA CNSS ACD CEDG
PDBID:2OYZ
HG QAP VGE SG EGN
PDBID:2P2E
DSN SLN DYKKDYN NI NTNN DNS
PDBID:2RLN
NLTKDR CKNG TGSS YP GNPY
PDBID:2XCZ

PDBID:2FLS

NC KTSC TVP NG

PDBID:2OSN

YIN GCYC QP RTAD MISASNSC

PDBID:2P8V

AR DPATK ATRN GA TPNM SQK

PDBID:2PAL

SFAG AADS DQDK SPSA DKDG

PDBID:2QIU

FVN TSI NY GERG

PDBID:2R39

DP DRNQ FRVAGE LNDV EPG SA

PDBID:2WQH

YPNN YPNN

PDBID:2XBD

WSD SSAW NGSQ NA SGST PNGS KNGS

PDBID:3C97

TN FD DIQMPVMD IDDD GAEL KPL

PDBID:3CR2

EGDKH SHFL DSDG

PDBID:3FWU

PEDLV DST FG LGG PLHC NG

PDBID:3KOX

KDG NG GS VTVVLPDV QKH DG AVGI

PDBID:3K63

RLKDT AA PNQ YQNE YPDG KKNN NPYG

DENG PG

PDBID:3K9I

AS FPNH DKLDE

PDBID:3KHQ

CSQ PR KRSS RGSQ GS QNI KCD

PDBID:3KU7

AATA RTL RTL LPYMEE

PDBID:3Q1D

CEEHEEE LSCE GAHKDC

PDBID:3R3K

QG

PDBID:3T3J

KPYTFEDY SG GDL PSS TGK SHDG LSSLAY

PDBID:4DP2

SG IGA NG

PDBID:3B7X

QRM ISGDRG APDA EHM LED TLG PPL

PPN

PDBID:3CR5

EGDKH SHFL DSDG

PDBID:3ETW

NTRFY

PDBID:3F8C

NGEME DEGR

PDBID:3FRV

PPK GQ

PDBID:3H9W

WQT HRDG DDSG

PDBID:3HQA

KH NNDE NNDE

PDBID:3IRC

TQHG DAP EKG TANP KE GEKA

PDBID:3L6W

PEKGDEK DWTQ IRSIN GTDSD GAKE TDDH KYDHT

DASVL RPNNAVF LPPK KRGT SVDEAVT PPFS LGKM

PEKGDEK DWTQ IRSIN KSDD GAKE TDDH KYDHT

DASVL RPNNAVF LPPK KRGT SVDEAVT PPFS LGKM

PDBID:3LHC

GS RTNG LNSVI DG GSS TAAG DG

PDBID:3NZ3

KVTN KVTG IDK DIP SDS

PDBID:3QD7

TGFL PLSQ REGL LLRQ DDKS FDD

PDBID:3RSD

DSST NLTKDR CKNG TGSS YP GNPY

PDBID:3T8T

KEGQ RGAKG RVP IESE

PDBID:3VXW

FKN AEKS IDKR ADL PPEKA ND PTAA

DKDG

PDBID:4BS7

LDN NF TQA NTDG GILQ SRWW DGRTPGS

NLCN SSD DGNGMNAW KGTD RGC

PDBID:4E1R

GADDGSLA PS PAG SEED AKG CSPH GM

PDBID:4DP4

ADDG PS PAG SEED AKG CSPH GM

PDBID:4EC2

SHG PAF PPNKQ PLS LNGE LRNG

PDBID:4GBN

TSI

PDBID:4IDL

QTGG APGK TTG GDFVKGR NANN DSL DGAR

PDBID:4JJD

DPN LYD ATLS TKG NHNGE KN

PDBID:4JZ3

TETD KKG NTEG SLTTG

PDBID:4KV4

DVEALGLH IKHP PPDH

PDBID:4MDQ

DEKQQH CSND VKE

PDBID:4QYW

PD MPEMN DPNA MG PF

PDBID:4XO2

MT MT

PDBID:5C4P

PD PG AEPP GAGE NQD GPGG

PDBID:5DKN

EGDKH SHFL DSDG FFE

PDBID:5EH4

EPE

PDBID:5HDG

DAL SPSG FD PC RG

PDBID:5NWX

TTG RTG GR LNDG YP

PDBID:5OC8

EK CSND VKE

PDBID:5UEV

DVEALGL IKHP PPDH

PDBID:5UVY

DVEALGLH IKHP PPDH

PDBID:6EVL

DPSH

PDBID:6PAZ

DDFDG DG DDFDG DG

PDBID:4EES

DPRLPDN ILG GPET TKSG DQKG

PDBID:4ETD

LDN NF TQA NTDG GILQ SRW DGRTPGS

LCN SSD DGNGMNAW KGTD RGC

PDBID:4HE7

ENY LAN DEKR YCE

PDBID:4HV2

LDN NF TQA NTDG GILQ SRW DGRTPGS

LCN SSD KGT RGC

PDBID:4IC9

VGVGGKS EPGD GLLN GLDT

PDBID:4IPF

DEKQQH CSND FGVQ VKE

PDBID:4MZ2

VNGT SAFGF SSDA EKS DDR SMNN RDL

PDBID:4N5T

GEGT EEV DKQRQH CHDD VK WAQSA

PDBID:4P15

GEG HSY

PDBID:4P7U

CEKPQE DEN DPKL HD LEDAAS PG SDE

PDBID:4P9E

CKRM NG PSGF NPFV IDGS

PDBID:4PE7

EGDKH SHFL DSDG

PDBID:4RLN

LDN NF TQA NTDG GILQ SRW DGRTPGS

LCN SSD DGNGMNAW KGTD RG GC CR

PDBID:4RUD

SSRTE PEGE HG SKC GAYN TDLC SRTE

PEGE HG SKC GAYN TD DL LC

CN NK

PDBID:4WO6

LDN NF TQA NTDG GILQ SRW DGRTPGS

LCN SSD DGNGMNAW KGT RGC

PDBID:5CPV

FAGVLN AADS GLTSK DQDK FKADA DSDG

PDBID:5EOZ

| | | | | | | | | | | | | | |
|---|---|---|---|---|---|---|---|---|---|---|---|---|---|
| AE | PA | NPG | VDK | IKDM | PEGA | KINE | VAG | PAG | PPAG | VDS | LSG | HN | PPAG |
| CTIH | GDSP | | | | | | RDG | | | | | | |
| PDBID:1B0T | | | | | | | PDBID:5E4P | | | | | | |
| IKCK | CPVD | PN | HPDEC | DGVKGK | | | LD | RG | NF | TQA | NTDG | GILQ | SRWWC |
| PDBID:1BFE | | | | | | | DGRTPGS | NLCN | SSD | NGMNAW | CKGTD | IRGC | |
| RGST | GEDGE | LAGG | RKG | NG | LRNA | GQ | PDBID:5EN9 | | | | | | |
| EAN | NSSG | | | | | | TSI | | | | | | |
| PDBID:1C0B | | | | | | | PDBID:5G25 | | | | | | |
| KET | DSST | NLTKDRC | CKNGQTN | KYPN | GNPY | | GG | AQDG | CPTAGD | EE | | | |
| PDBID:1C76 | | | | | | | PDBID:5UEP | | | | | | |
| DGKG | KPG | TAYKE | DPSA | KNKK | TEKG | LSEHIKNP | DVEALGL | IKHP | PPDH | | | | |
| PDBID:1CUO | | | | | | | PDBID:5UVZ | | | | | | |
| STR | PASCA | PKTGMG | GGG | SKDE | YPGHFSMM | | DVEALGLH | IKHP | PPDH | | | | |
| PDBID:1DDV | | | | | | | PDBID:5VR3 | | | | | | |
| DPNTK | STRN | GS | TPNM | SQK | | | PAIP | EHNP | | | | | |
| PDBID:1E5B | | | | | | | PDBID:5XWE | | | | | | |
| WSD | SSA | NGSQ | SGST | NGSGN | KNGS | | IDSC | PEGE | YT | PKTE | YT | IDSC | PEGE |
| PDBID:1EFQ | | | | | | | WKI | PKTE | | | | | |
| PD | SLG | SQS | SSNS | KPGQ | WAS | ESGVPDR | PDBID:6FGT | | | | | | |
| GT | SS | YYSTPY | | | | | SMSV | HEDAWPFL | NLKLVPG | IKKP | EDDS | | |
| PDBID:1EN2 | | | | | | | PDBID:6H0L | | | | | | |
| SQGG | IWG | CGRT | RSDH | GN | GQD | VHG | LDN | NF | TQA | NTDG | GILQ | SRW | DGRTPGS |
| PDBID:1E0T | | | | | | | LCN | SSD | DGNGMNAW | KGTD | RGC | | |
| PASVP | CFNL | PLQR | SGKCPQK | KLA | DPKK | | PDBID:1A18 | | | | | | |
| PDBID:1FA0 | | | | | | | CDAFV | GD | STFK | TADD | GG | DG | GD |
| LG | GGLVK | RN | KDQMSPE | YSQERV | PF | | KG | | | | | | |
| PDBID:1FAZ | | | | | | | PDBID:1B6E | | | | | | |
| DLCTQA | NPFGFP | | | | | | QEK | RC | WENG | SQYLFPS | NTK | NPNG | CED |
| PDBID:1FDA | | | | | | | PDBID:1C9V | | | | | | |
| IKCK | CPVD | PN | HPDEC | CPAQ | PEDM | LPD | DSST | NLTKDR | CKNG | TGSS | YP | GNPY | |
| WDGVKGK | | | | | | | PDBID:1CS3 | | | | | | |
| PDBID:1FE3 | | | | | | | GTLC | DSQ | SQ | DFL | | | |
| VEAFC | GD | STFK | LG | TADD | DGDK | DG | PDBID:1DCF | | | | | | |
| DG | GD | | | | | | FTGL | SHEHK | VEN | KPV | | | |
| PDBID:1FER | | | | | | | PDBID:1EMN | | | | | | |
| IKCK | CPVD | PN | HPDEC | EDEVPEDM | DGVKGK | | DG | PFG | GN | NPCGNG | IG | TCEEGF | GPMM |
| PDBID:1FEV | | | | | | | PDBID:1E0Q | | | | | | |
| NLTKDR | CKNG | TGSS | YP | GNPY | | | GPSE | STL | | | | | |
| PDBID:1F0W | | | | | | | PDBID:1E0W | | | | | | |
| MTFITKT | IESG | RSMG | | | | | DSST | NLTKDR | CKNG | TGSS | YP | GNPY | |

PDBID:1GD6
SR TSK NRNG GLFQ DRYW KDCN TDD
RFDAW HCQGS
PDBID:1GHJ
PESIA WHKKPGE KRD DKV NEGD LSG
PDBID:1GKH
AQ SRQG NEYP DEGQ GQFG DRL
PDBID:1GV2
GPKR LKGR GNR LPGR
PDBID:1GXT
VQGV DGDG PPLA SQL
PDBID:1H2P
RN VPGG FG NTG ST ISGSS IDN
QGERDH YR NKGF EH NNDE
PDBID:1IDI
TAT NLCY CDAF SRGK KPYE TDKC
PDBID:1J7Z
NLTKDR CKNG TGSS YP GNPY
PDBID:1JPO
LDN NF TQA NTDG GILQ SRWW DGRTPGS
NLCN SSD KGTD RGC
PDBID:1KM8
DN GG TGV PRPC NQ GR
PDBID:1KSM
KEGDPNQL DKNG
PDBID:1KTH
PNTK NENK
PDBID:1LOZ
MDG GY TRA AGDR GTFQ SRYW DGKTPGA
NACH QDN DPQ QNR VQGC
PDBID:1MH7
VPAR GCY GG QG KGRN NI
PDBID:1O4K
EEW KI NAENPRG SETTKGA NAK LDSG SR
ADGLCHR
PDBID:1O4L
EEW KI NAENPRG SETTKG NAK LDSG SR
ADGLCHR
PDBID:1O4Q
EEW KI NAENPRG SETTKG NAK LDSG SR
ADGLCHR
PDBID:1RTU
PQ GG YP SEDI VYNG RD TNTG

PDBID:1EQV
GD KG LL ETK PKK AKK DKYG
DG AP
PDBID:1FIK
YKDSP VPGK GG LLQDGEF TGGA TDK GV
PDBID:1FKD
DGRT KRG EDG GK SVG ATG PGI
PPH
PDBID:1FKF
DGRT KRG EDG GK SVG ATG PGI
PPH
PDBID:1HEY
DKEL TGKS NN WNMPNMD DGAMSAL KPF
PDBID:1HME
DPNA HPGL
PDBID:1IOS
LDN NF TQA NTDG GILQ SRWW DGRTPGS
NLCN SSD DGN KGTD RGC
PDBID:1JPE
QH EFY DAG
PDBID:1KM9
DN GG TGVI PRPC NQ GR
PDBID:1KXX
LNN NF TQA NTDG GILQ SRWW DGRTPGS
NLCN SSD DGN KGTD RGC
PDBID:1LXI
WQDW AFPL NPETV DDSS
PDBID:1MG6
GCNC NK HL
PDBID:1O2E
IPSS NN GCY NPYTN NN SSEN NLDMKN
PDBID:1O45
EEW KI NAENPRG SETTKG NAK LDSG SR
ADGLCHR
PDBID:1O4G
SIQAEEW KI NAENPRG SETTKGA NAK LDSG SR
ADGLCHR
PDBID:1Q8B
DLNEI EEG
PDBID:1R26
MRAR AVWC FPTV ADNN QLP SG GAN
PDBID:1RLK
KDLD GYTQVE

| | | | | | | |
|---|---|---|---|---|---|---|
| HTGA | SYDG | | | | | |
| PDBID:1RZY | | | | | | |
| KEIPA | DDQ | DISP | AP | KKH | LKKG | QSVY |
| NWPP | | | | | | |
| PDBID:1TUW | | | | | | |
| HD | RSH | AQDW | | | | |
| PDBID:1U3J | | | | | | |
| DRT | VTG | KD | EV | RQF | YKDVNI | KSDL |
| PDBID:1UIC | | | | | | |
| LDN | NF | TQA | NTDG | GILQ | SRWW | DGRTPGS |
| NLCN | SSD | DGN | KGT | RGC | | |
| PDBID:1VHF | | | | | | |
| RLIA | KG | YETP | LTEY | | | |
| PDBID:1ZIA | | | | | | |
| KD | PA | APG | PTDK | IKGM | PDGA | KINE |
| CTPH | GDAP | | | | | |
| PDBID:1ZJ7 | | | | | | |
| LWQRP | GG | DTGA | IG | CG | | |
| PDBID:2B8G | | | | | | |
| KAG | EKG | MK | DRS | KEG | NEG | SNST |
| PDBID:2BRF | | | | | | |
| PPGE | PSDGQ | PLTQ | DRKCSRTQ | PETR | GQ | KPG |
| VG | NG | | | | | |
| PDBID:2BZY | | | | | | |
| DKT | EVG | NING | NG | DKT | EVG | NING |
| NG | PQ | QN | | | | |
| PDBID:2CW4 | | | | | | |
| TDRAP | AQ | GG | IP | APDG | DE | TPPY |
| PDBID:2HB4 | | | | | | |
| LWKRP | GG | DTGA | IG | CG | GPTPVN | |
| PDBID:2J5A | | | | | | |
| KPTL | QK | DED | | | | |
| PDBID:2JK7 | | | | | | |
| PRN | FGTW | GD | FHCG | KPSE | YPGC | |
| PDBID:2NUH | | | | | | |
| RL | WQGK | YRLP | | | | |
| PDBID:2OPY | | | | | | |
| PRN | GTW | GD | FHCG | KPSE | PGC | |
| PDBID:2PPI | | | | | | |
| PDBID:1SDZ | | | | | | |
| FTDW | LDWL | GD | FFCG | EQED | PIN | |
| PDBID:1SNQ | | | | | | |
| GD | KG | LL | ETVEKY | AKK | DKYG | DG |
| KG | | | | | | |
| PDBID:1SSC | | | | | | |
| DSST | NLTKDR | CKNG | TGSS | YP | | |
| PDBID:1TAY | | | | | | |
| MDG | GY | TRA | AGDR | GIFQ | SRAW | DGKTPGA |
| NACH | QDN | DPQ | QNR | VQGC | | |
| PDBID:1TXA | | | | | | |
| PDAT | KPGV | DNC | TWKRK | | | |
| PDBID:1U68 | | | | | | |
| YNR | ENP | | | | | |
| PDBID:1U9R | | | | | | |
| GD | KG | LL | ETVEKY | AKK | DKYG | DG |
| KG | | | | | | |
| PDBID:1UID | | | | | | |
| LDN | NF | TQA | NTDG | GILQ | SRWW | DGRTPGS |
| NLCN | SSD | DGNGMNAW | KGTD | RGC | | |
| PDBID:1YEB | | | | | | |
| QQCH | EEGG | NKVGP | LHG | SGQVKGY | TN | IPGT |
| PDBID:208L | | | | | | |
| MDG | GY | TRA | AGDR | GIFQ | SRYW | DGKTPGA |
| NAAH | QDN | DPQ | QNR | QGC | | |
| PDBID:2D4M | | | | | | |
| ATPGS | PPN | DNDY | VNN | SQG | | |
| PDBID:2F03 | | | | | | |
| PIN | HPNNIR | LENTIYAN | PDDYPLKP | QKP | TH | YSNG |
| GDDYNPSL | | | | | | |
| PDBID:2FTB | | | | | | |
| NGD | PN | IG | SMGG | GG | TDQ | GN |
| GG | | | | | | |
| PDBID:2H8A | | | | | | |
| NKV | NDL | VPF | SLSG | QPNR | | |
| PDBID:2JP8 | | | | | | |
| RVYI | | | | | | |
| PDBID:2MSS | | | | | | |
| SVNT | VD | DKTTNRHR | FES | NN | KA | |

DG EGI

PDBID:2V6H

TVGG GKW QH RASK TDA TKDK

PDBID:3AOV

KDG VLG LPD KGER NAKTQ

PDBID:3BI9

LGQ PNSKCNA DGTR STK KVQFG SNT GWFN

AL LV

PDBID:3ENU

KNF PRL PFG WENK GPR HNY DAGA

ANL FDNF

PDBID:3EZM

GS RTNG LNSV DG NFIET GSS TRAQ

DG

PDBID:3FAJ

KKGD

PDBID:3GLW

RNF TR GQ

PDBID:3GM3

RTDD AQL AVT

PDBID:3GQU

PSS LGAHG PWW IAE

PDBID:3I3Z

TSI FVNQ

PDBID:3I7W

DSSTS NLTKDR KP CKNG TGSS YP GNPY

PDBID:3I7X

DSSTS NLTKDR KP CKNG TGSS YP GNPY

PDBID:3IE9

ADGA KM ETP KVG VAGVLG KKE TP

PDBID:3INC

TSI

PDBID:3J4G

LDNYRGY NF TQA NTDGS GILQ SRWW TPGS

NLCN SSD KGT IRGC

PDBID:3JZR

KDT DEKQQH CSND VKE

PDBID:3KVT

GG KIPAT TEGMLN PVLN PTDVC

PDBID:3ONH

PQD ASNQ YD EDLNDR GNG EEGDT DELPCNT

PDBID:3S9K

NLETYEW SI KEG SRTPGT KAIISEN DSPK FDS

PDBID:2OQK

EEG NG FDG NPG DF GEIPETT

PDBID:2OUB

SS GCY GWG NG LY

PDBID:2P61

SPT KL EWQTI

PDBID:2PNE

CPG GPG NPGC GT GTPK

PDBID:2PVT

GCY GG NG LY

PDBID:2QDB

GN KGQ LL AKK DKYG DG KG

PDBID:2QHW

GCY WG NG LY

PDBID:2REA

DGR YNPD EGQ RMVLGRT

PDBID:2REY

MEYN MKGG EPG ND FENM

PDBID:2UUX

PIGW MGNC

PDBID:2VSL

PRN FGTW GD FHCG KPSE PGC

PDBID:2W51

RPGDC TFS TDDA DSQICEL KKL GETCKGC

PDBID:3BKS

DRSKR EG FNDL ELAV TADYP KETS

PDBID:3CQ1

DPELG VVNLG PP PLHD LPGV FEPP RLL

PDBID:3DML

QPGC QRD PPGL LARP FTP GD

PDBID:3G7C

WIREYP REES

PDBID:3KLU

LSFFPGQT PISKRF DKEG TTEL PTFKA

PDBID:3L1M

GK

PDBID:3LLH

EPN GD GQP GD PS

PDBID:3O70

LYFQGL FAGR NECH KS VPEV

PDBID:3P2X

CTS

PDBID:3PAZ

GGGLVTR SPGI

PDBID:3UAF

CPTD KK DRNG GP HRDSN PS LDR

PDBID:3WVT

PQCKEE GT SPG GELT APR LSKC ST

DQG GD

PDBID:4DPO

ADDGSLA PS PAG SEED AKG CSPH GM

PDBID:4F68

FGFKGV

PDBID:4FE6

LWQRP GG DTGA IG CG PTPVN

PDBID:4GQV

KPTT DEDW EKT DSDG

PDBID:4K59

EGE KADY GN PRG

PDBID:4KGT

GG DG GGL DG

PDBID:4LFQ

RKTCG

PDBID:4LTT

KPF PD DD

PDBID:4NEJ

KKA

PDBID:4OXW

EGKSK GG PELV IYSAL

PDBID:4TQN

VKNP YQE RKTS

PDBID:5B79

CDFCL SKINKKTG CSDCG KTYRW CIECK NICG SEN

CDDCD LTPSMSEP ASIY

PDBID:5BPO

TSI TSI

PDBID:5PAL

DPGT LKGK DKDQ SAHG DSDH

PDBID:5UVS

DVEALGL IKHP PPDH

PDBID:6BBK

AE PA NPG VDK IKDM PEGA KINE

CTPH GDSP

PDBID:3RNT

CD GS GSNS NYEGFDF LSSG ENN HTGASGNN

PDBID:3UB2

SRWSKD GAI QALS APGAE SGL GRG

PDBID:3UB3

SRWSKD SEE GAI QALS PGAEG SGL GRG

PDBID:3UMD

EHVAFGS GDG VIG DSPEF DYQM SGD

PDBID:3UME

AFGS GDG VIG DYQM SGD

PDBID:3V1G

TSI GRG TSI GRG

PDBID:3WUN

LDN NF TQA NTDG GILQ SRW DGRTPGS

LCN SSD DGNGMNAW KGT IRGC

PDBID:4AYA

VPSIPQNK HLKPSF VPSIPQNK

PDBID:4CGQ

QPQN AEHV RMEE

PDBID:4DH4

VAA GG GG GD

PDBID:4ETA

LDN NF TQA NTDG GILQ SRW DGRTPGS

LCN SSD DGNGMNAW KGTD RGC

PDBID:4ETB

LDN NF TQA NTDG GILQ SRW DGRTPGS

LCN SSD DGNGMNAW KGT RGC

PDBID:4ETE

LDN NF TQA NTDG GILQ SRW DGRTPGS

LCN SSD DGNGMNAW KGTD RGC

PDBID:4EWW

TSI

PDBID:4EWZ

TSI

PDBID:4F6D

VN FGFKGV

SATPKDT STL NQDG KNDG

PDBID:6CEC

CPH PAAG DVTQ CGDCG IQE LSCY GRYING

YIDL YYCQ

PDBID:6CEF

PWCPH PAAG DVTQ CGDCG IQE LSCY GRYING

YIDL YYCQ

PDBID:6EVM

DPSH

PDBID:6FH6

SMSV HEDAWPFL NLKLVPG IKKP EDDS

PDBID:6IFH

MKIPGMD TPDV AYGE KPF

PDBID:1AB0

GDAFV GD STHK TADD GG DG GD

KG

PDBID:1ACD

GDAFV GD STHK TADD GG DG GD

KG

PDBID:1B1J

QDNS DAKPQGR TSPC ENKN REN WPPC ENGL

QSIFRR

PDBID:1BM2

HPW KI HDG SESAPGD GN DGAG LWV

RNQ

PDBID:1BQK

KD PA APG PTDK IKGM PDGA KINE

CTPH GDAP

PDBID:1D1M

NADG DG NADG DG

PDBID:1DDW

DPNTK STRN GS SQK SRAN

PDBID:1E6L

DKEL WNMPNMD SAL AEA KPF

PDBID:1FDD

IKCK EVCPVD PN HPDEC PEDM DGVKGK

PDBID:1FE5

GCY EP DTAD MISSSTNC

PDBID:1FF2

IKCK PN HPDEC PEDM DGVKG

PDBID:4GSP

CD GS GSNS NYEG LSSG ENN HTGASGNN

PDBID:4HB6

KAKNG VPD GKGC

PDBID:4JHB

GS NKE CTG KE TASW RQF YEDLEI

RLTDG

PDBID:4LJN

SFCL KEQNREK CADCG CIECK SSCR ADN CDSCD

CDPP PKGM KK KG

PDBID:4P2P

IPGS NN GCY LDSC NT NSKN NLDTKKY

PDBID:4RLM

LDN NF TQA NTDG GILQ SRW DGRTPGS

LCN SSD DGNGMNAW KGT RG GC CR

PDBID:4YM8

LDN NF TQA NTDG GILQ SRW DGRTPGS

LCN SSD NGMNAW KGT RGC

PDBID:4YOP

LDN NF TQA NTDG GILQ SRW DGRTPGS

LCN SSD DGNGMNAW KGTD RGC

PDBID:5C09

TSI

PDBID:5DM9

LDN NF TQA NTDG GILQ SRW DGRTPGS

LCN SSD NGMNAW KGTD IRGC

PDBID:5HBP

IPE HE APA ED

PDBID:5HMB

PD PHG GDR GAT DAAG

PDBID:5HQI

TSI FVNQ

PDBID:5LAW

EKQQH CSND VKE

PDBID:5LN2

SQIP SVG DEKQQH CSND VKE

PDBID:6CE6

CPH PAAG DVTQ CGDCG IQE LSCY GRYING

YIDL YYCQ

PDBID:6DXH

PDBID:1FLQ

LDN NF TQA NTDG GILQ SRWW DGRTPGS

NLCN SSD DGNGMNAW KATD RGC

PDBID:1FLY

LDN NF TQA NTDG GILQ SRWW DGRTPGS

NLCN SSD DAN KGTD RGC

PDBID:1GOD

KDKT

PDBID:1HEH

VRG WAD SGSSSW NGGQ PNGS KNGS

PDBID:1HRC

AQCH EKGG HKTGP LHGLFG TGQAPGF IPGT

PDBID:1J73

GERG

PDBID:1JZA

KSTG AKNQG YAF GLPEST PN KSTG AKNQG

YAF GLPEST PN

PDBID:1K1Z

AQDKKR ELGL GIPP PGAF NPG AEAEH TATN

PCNR

PDBID:1M1S

PP PATG SNNEH PV DAKG PAEE APFKAG

PDBID:1O41

EEW KI NAENPRG SETTKGA NAK LDSG SR

ADGLCHR

PDBID:1O47

EEW KI NAENPRG SETTKGA NAK LDSG SR

ADGLCHR

PDBID:1O4R

SIQAEEW KI NAENPRG SETTKGA NAK LDSG SR

ADGLCHR

PDBID:1OR5

SALT EDDSV RYG

PDBID:1PAL

SFAG AADS DQDK FSPSA DKDG

PDBID:1Q4R

KDGV HQGY

PDBID:1QL2

IDT IDT IDT

PDBID:1T00

HMAG TDD LKN AAWC GDKI IDEN SIP

GG

PDBID:1UIA

SKHAFSL CSKC REN LTDG YFDG PD YDED

DMLKM

PDBID:6INS

GERG TP PKG TSI

PDBID:6JQ4

DPEN GKRK

PDBID:1A0B

SPDL

PDBID:1AF5

NKE PNQSYKFK TQR GS LKLKQ AALN

PDBID:1ALB

CDAFV GD STFK LG TADD GG DG

GD KG

PDBID:1B1I

DAKPQGR TSPCKD ENKN REN WPPC NG

PDBID:1BAS

KNGG HPDG EKSDPHI RG VSAN KEDG TDEC

ESNN RKYTSW KRTG GPGQKAI

PDBID:1CYI

AACH NSVMPEKT LDGG GA WADRL

PDBID:1DKJ

LDN NF SQA NTDG GVLQ SRWW DGKTPGS

NLCN SSD KGT RGC

PDBID:1EV3

TSI TSI

PDBID:1HE7

GQPA NG TSF AANE QP PF NPFE

PDBID:1HQ8

PNNW RN IYS IPANG EDG SYNQL IPKG

SSF DCAN

PDBID:1I0V

CD GS GSNS NYEG LSSG ENN HTGASGNN

PDBID:1I3Z

LPY GCL VDG ESVPG KK EKHGY DAHT

PN

PDBID:1I0R

LDN NF TQA NTDG GIFQ SRWW DGRTPGS

NLCN SSD DGN KGTD RGC

PDBID:1IRW

EKGG HKVGP LHGIFG SGQAEGY IPGT

PDBID:1JDL

CMACH GPDA NLVGP LTGVID AGTAPGF PD

PDBID:1JKD

DN NF TQA NTDG GILQ SRWW DGRTPGS
NLCN SSD GN KGT RGC
PDBID:1UWM
EHNG KPG AYEPNPAT DG
PDBID:1WAQ
WDDW PL EFPL DPEST DSAN
PDBID:1WHI
QE GSGR NIG TKRG RPDG RDDK AP
PDBID:1WY9
SND NGNG SEE
PDBID:1YGT
GGN KP NNDTD NKT
PDBID:1Z21
SNPP
PDBID:2A9I
TPST PQE KPSG RYN LQTGL
PDBID:2AZ8
LWKRP GG DTGA LPG IG CG GPT
PDBID:2BQQ
NG RYF SSDRFR NI LPQG IDG EEGE
DDV NPNW
PDBID:2COQ
PR ETGE DTA LGS GR RDL AFN
PDBID:2EH9
ELG GK GEEP
PDBID:2HWV
GD AY RG IG FGD ED PS
PTY RGV
PDBID:2JIN
DYLV PS LG KENG QEG NG LKNL
GY
PDBID:2O0P
RGQAN AEPGED DADG
PDBID:2O0Q
RGQAN GED DADG
PDBID:2OAI
REDG DGTL NNYH LAG HVG AG GA
PDBID:2OJR
DTNN EGDE DTNN EGDELLA TLTG EPSD AG
EDGR LSDYN QKES LR RG
PDBID:2ON8
GKTL
PDBID:2P1X

MDG GY TRA AGDR GIFQ SRYW DGKTPGA
NACH QDN DPQ QNR VQGC
PDBID:1KH0
FANG NG FANG NG
PDBID:1KH8
DSSTS NLTKDR CKNG TGSS YP
PDBID:1KXI
LPFI PEGK KKFPL DNC SAL TDKC LPFI
PEGK KKF DNC SAL TDKC
PDBID:1L5D
LDVR GPGGK GPS DE GT NYLP GSG
AAAG SPDG DP PA AC CD DP
PE EA FS AF
PDBID:1LVE
PD SLG SQS SSNS KPGQ WAS ESGVPDR
GT SS QAEDV STPY
PDBID:1MB3
LPEI DDDLAHI KPI
PDBID:1NEH
PAN AADD QD QF QLFPGK VNG CASW
PDBID:1O3X
NKLI SQG TEDN
PDBID:1O5J
KG YETP LTEY
PDBID:1PZC
AE PA NPG VDK IKDM PEGA KINE
CTPH GDSP
PDBID:1REX
MDG GY TRA AGDR GIFQ SRYW DGKTPGA
NACH QDN DPQ QNR VQGC
PDBID:1RFP
LDN NF TQA NTDG GILQ SRWW DGRTPGS
NLCN SSD DGN KGTD RGC
PDBID:1RS2
NVIDT YNR ENP
PDBID:1SNP
GD KG LL ETVEKY AKK DKYG DG
KG
PDBID:1T2J
QPGG APGK SGSG ADSVKGR NSL DWYGMD
PDBID:1TCY
MDG GY TRA AGDR GIFQ SRFW DGKTPGA
NACH QDN DPQ QNR VQGC

GG ENG KQG GD DRSG

PDBID:2Q1M

GPLPSK SEPP DW NANY NK NKSK HVG

NSEHQ

PDBID:2X44

PAVV SSRG AT DS MMGN LDDS SGN

PDBID:2YHN

CMVCCEE PCG CPVCR CMVCC PCG CPVCR

PDBID:2Z44

PIQE LPE

PDBID:2ZHH

NSGNQ SRSDC

PDBID:3A7L

PAEL SKEH EADG EVG SAG VKA PVS

DSP EPYAGG SD

PDBID:3BOV

APKE VG ELE SLQSERA PS GA

PDBID:3BY5

RKGA KADE TFSQASL GAGA GD

PDBID:3DP5

ASWS AV NTVHPEKT RNPGPGM GEAMIP

PDBID:3EAZ

MPW KI YPPETG TNYPG DG AS IDEEV

ADGLCTR

PDBID:3ERS

HENADKL VGQK LVPYYS MG CNL RG TDDG

PERM PAG

PDBID:3GBQ

ADD KRG QN NG

PDBID:3GK2

EGDKH SHFL DSDG

PDBID:3GK4

EGDKH SHFL DSDG

PDBID:3GSP

CD GS GSNS NYEG LSSG ENN HTGASGNN

PDBID:3HJX

RPMDEYS TT

PDBID:3M4T

GLFC ED SVA SFI IVRV

PDBID:3OZZ

AK HY LSEHPGG AG

PDBID:3UNN

AH MPDCS WD CGSLNG RPP RDQ AD

PDBID:1TVQ

GD PR LG TMDG NG EK GN

GG

PDBID:1UW3

LGGY FGN

PDBID:1VED

LDN NF TQA NTDG GILQ SRWW DGRTPGS

NLCN SSD NGMNAW KGT IRGC

PDBID:1Z3L

TAAA NLTKDR CKNG TGSS YP GNPY

PDBID:2A9Q

PD KPF

PDBID:2APB

AA PR VTG DTGH AG GDIPDG QE

ES GGGT

PDBID:2B29

SEG DTNI GN DGL EQLSSN LKDG SAEAVG

PDBID:2FKE

DGRT EDG GK SVG ATG PGI PPH

PDBID:2FKL

DKC RMDV ID PDKC RMDV ID

PDBID:2FMB

LEKR ND DTGADTS LKYR VG KG DI

PDBID:2I9V

AFGS DDG DGDG DPKQVIG DSPEF DQQM ALSG

PDBID:2IGP

PLND QAP CPSY DTK VGAPH

PDBID:2PK7

CPICK SADK DG CPICK SADK DG

PDBID:2PWS

GCY WG NG LY

PDBID:2QAS

DLMQ APGG PEPH TKAAGV YPD QHQ GET

GG FSA

PDBID:2QVT

SPTA SGL NEQM QAEE GPKLNL SNMDQH SGAT

NDIPNF

PDBID:2VKS

AAA AV EA DV DGPG TD SGA

PDBID:3AOS

KDG VLG LPD GE NAKTQ

PDBID:3BIA

LG PNSKCNA DGTR STK SN GWFN

PDBID:3V19

HSI

PDBID:3VYA

ED DGN GGM RDE RKV RN IARFKWA

PDBID:4AHI

DAKPQGR GLTSPCID ENKNG REN GGS WPPC NG

PDBID:4AQI

GRDG DKNE

PDBID:4AQJ

RRDG NYLA DKNE

PDBID:4EXO

TSI

PDBID:4LYO

LDN NF TQA NTDG GILQ SRWW DGRTPGS

NLCN SSD GNGMNAW KGTD IRGC

PDBID:4UNG

TSI TSI

PDBID:5AFG

DAAQQH CSND VKE

PDBID:5B1G

LDN NF TQA NTDG GILQ SRW DGRTPGS

LCN SSD NGMNAW KGT RGC

PDBID:5BMH

GKTL

PDBID:5C6X

PD PG AEPP GAGE GPGG

PDBID:5CUL

SS PTH RRGETPLP NVD

PDBID:5D53

TSI FVNQ

PDBID:5FD1

IKCK CPVD PN HPDEC CPAQ DGVKGK

PDBID:5KAZ

LPY HGRL VDG SESIPG KN EKHGY AEGS

PS

PDBID:5NGN

DSCSEYC CCP DSCSEYC DGQ CCP

PDBID:5OC4

PAQI AE

PDBID:5PAZ

AE PA NPG VDK IKDM PEGA KINE

CTAH GDSP SA

PDBID:3C4S

FPG NVDDTYYR DG GN FPG NVDDTYYR DG

GN

PDBID:3C5K

PWCPH PAAG DVTQ CGDCG IQE LSCY GRYING

YIDL YYCQ GEDM

PDBID:3C7I

HPW KI HDG ESAPG GN DGAG LWV

RST SRNQ

PDBID:3DJN

NHK PSE PY GEDG

PDBID:3FFY

NAT PNEK PQKKGR VDE GPER KI

PDBID:3GKY

HSI HSI GERG

PDBID:3HQB

YKH NNDE NNDE

PDBID:3IN2

AECS DQM NTN DKSCKQ PKNVMG TAAD GSG

FPGHSALL

PDBID:3LRO

DPD EKTD GDDT ND DD

PDBID:3LVE

PD SLG SQS SSNS KPGQ WAS ESGVPDR

GT SS QAEDV YYSTPY

PDBID:3MF8

QDRLT PAGN GG VEYG

PDBID:3MYA

PRGVDPSR LVNT PRGV PS

PDBID:3Q7Y

TDTG NPDG DRSDPGI DGNG TDTG NPDG DRSDPGI

DGNG TDTG NPDG DRSDPGI

PDBID:3RHE

KN PIES PT VKTG IEPKA SNE QDF

DPDE

PDBID:3STM

KGKDI GK GS TMTG EGDN KN GD

GD

PDBID:3SUL

NADQ VACS GPNG SGTG AGNG NGQ

PDBID:3T1X

DRKG ALWA PPP LLG GERM GEHA DETA

PDBID:5TAB

GPLGSEV RCICE NDF CEECQ CYVC

PDBID:5UET

DVEALGL IKHP PPDH

PDBID:5XUK

KRFK KDK GKELS SPKN AG

PDBID:5ZND

RDDVA PDCDDW DPHLCD

PDBID:6CEE

CPH PAAG DVTQ CGDCG IQE LSCY GRYING

YIDL YYCQ

PDBID:6EKB

CANCEGEG CSQCKGG HFNG KAG CWLCRGK CGDCNGA

PDBID:6I3S

EKQQH CSND VKE

PDBID:6QQJ

DPRLPDN ILG GPET TKSG DQKG

PDBID:6RNT

CD GS GSNS NYEGFDF LSSG ENN HTGASGNN

PDBID:1B2O

TVC NPGT PDDW CPLCA TVC NPGT PDDW

CPLCA

PDBID:1BKF

DGRT KRG EDG NK GK SVG ATG

PGI PPH

PDBID:1BPQ

IPSS NN GCY NPYTN NN SSEN NLDKKN

PDBID:1COC

KET DSST NLTKDRC CKNGQTN KYPN GNPY

PDBID:1C9H

DGRT KKG QNG SRDRN GK SLG ATG

PGV PPN

PDBID:1DMM

GD DPFGQ SHNG NG DEHG

PDBID:1DMQ

DPFGQ SHNG NG DEHG

PDBID:1DYZ

AQC NDAM NVK DKSCK AKVAMG GGG FPGHWAMM

PDBID:1DZO

PDBID:3WW5

LDN NF TQA ETDG GILQ SRW DGRTPGS

LCN SSD DGNGMNAW KGTD RGC

PDBID:3ZEK

LDN NF TQA NTDG GILQ SRWW DGRTPGS

NLCN SSD DGNGMNAW KGTD RGC

PDBID:4AHG

TSPC ENKN REN GG WPPC NG

PDBID:4ET9

LDN NF TQA NTDG GILQ SRW DGRTPGS

LCN SSD DGNGMNAW KGTD IRGC

PDBID:4GFY

SS GCY GWG VNGA LYPDFLCK

PDBID:4HMB

GCY WG NG LY

PDBID:4HP9

FQSR ARVENC MQ CTC GPRT RKDG NEDG

PDBID:4HRS

GR VVGE DDDG DG PEDQ VEV RLIK

PDBID:4IAS

LDN NF TQA NTDG GILQ SRW DGRTPGS

LCN SSD DGN KGTD RGC

PDBID:4KUO

DPSQPDN VLG GPDT RKDG DPEG

PDBID:4LFS

SCI RKTCG

PDBID:4ML2

TLPL KG PD TDK

PDBID:4P9V

HPW KI HDG ESAPG GN DGAG LWV

SRNQ

PDBID:4PTA

NW DTEG KRPN

PDBID:4XC4

TSI

PDBID:5AEF

GVVI GVVI

PDBID:5CB9

PD PA PG AEPP GAGE NDTACCY GPGG

| | | | | | | | | | | | | | |
|---|---|---|---|---|---|---|---|---|---|---|---|---|---|
| | | | | | | | | | | | | A | |
| AQC | NDAM | NVK | DKSCK | AKVAMG | GGG | FPGHWAMM | PDBID:5J6X | | | | | | |
| PDBID:1E6M | | | | | | | DN | KQ | DN | | | | |
| SDKEL | DDF | WNMPNMD | DGAMSAL | KPF | | | PDBID:5K2L | | | | | | |
| PDBID:1E97 | | | | | | | QPGD | | | | | | |
| DPFGQ | SHNG | NG | DEHG | | | | PDBID:5O3A | | | | | | |
| PDBID:1EA2 | | | | | | | IKHP | PPDH | | | | | |
| DPFGQ | SHNG | NG | DEHG | | | | PDBID:5UER | | | | | | |
| PDBID:1ED1 | | | | | | | DVEALG | IKHP | PPDH | | | | |
| LRPGG | VPTG | TGTA | | | | | PDBID:5VEA | | | | | | |
| PDBID:1FAV | | | | | | | LWKRP | GG | DTGADD | IG | CG | PTPV | |
| EL | | | | | | | PDBID:5WPB | | | | | | |
| PDBID:1G2S | | | | | | | CPH | PAAG | DVTQ | CGDCG | IQE | LSCY | GRYING |
| SDISKT | SNSCSQR | KRG | HPRK | TPK | | | YIDL | YYCQ | | | | | |
| PDBID:1HKF | | | | | | | PDBID:5WRB | | | | | | |
| VAGQ | TGSL | AL | KPR | TSR | TDL | RPSDN | LDN | NF | TQA | NTDG | GILQ | SRW | DGRTPGS |
| PDBID:1IOC | | | | | | | LCN | SSD | DGNGMNAW | KGT | RGC | | |
| NSS | MKD | RQ | NASG | NSS | MKD | RQ | PDBID:6FH7 | | | | | | |
| NASG | NNI | | | | | | SMSV | HEDAWPFL | NLKLVPG | IKKP | EDDS | | |
| PDBID:1IOT | | | | | | | PDBID:6MHN | | | | | | |
| LDN | NF | TQA | NTDG | GILQ | SRWW | DGRTPGS | AFGS | GDG | DPKQVIG | DSPEF | DYQM | ALSGD | |
| NLCN | SSD | DGN | KGTD | RGC | | | PDBID:8PAZ | | | | | | |
| PDBID:1J4H | | | | | | | AE | PA | NPG | VDK | IKDM | PEGA | KINE |
| DGRT | KRG | EDG | GKQEVI | SVG | ATG | PGI | CTPH | GDSP | | | | | |
| PPH | | | | | | | PDBID:134L | | | | | | |
| PDBID:1JC7 | | | | | | | MDG | GY | TRA | AGDR | GIFQ | SRYW | DGKTPGA |
| AN | PVHH | GGN | GDPLICDN | STG | WPAHRNE | | NACH | QDN | DPQ | QNR | VQGC | | |
| PDBID:1JHC | | | | | | | PDBID:1A6F | | | | | | |
| AAGEPLLA | DPSLFKPN | MA | MDG | QDVRNG | DD | GN | NRQ | QPEN | SKKIG | ER | KEK | SQL | |
| NSEF | LRQQ | DWLV | | | | | PDBID:1BEL | | | | | | |
| PDBID:1JON | | | | | | | DSSTS | NLTKDRC | CKNG | TGSS | YP | GNPY | |
| DR | LS | IN | SP | GIVK | NKDT | | PDBID:1BGI | | | | | | |
| PDBID:1KC2 | | | | | | | LDN | NF | TQA | NTDG | GILQ | SRWW | DGRTPGS |
| EEW | KI | NPENPRG | SETTKGA | LNV | LDSG | SR | NLCN | SSD | NGMNAW | KGT | RGC | | |
| ADGLCHR | | | | | | | PDBID:1BOO | | | | | | |
| PDBID:1KP4 | | | | | | | PVGI | TSSHCPRE | KLD | KKT | TPK | | |
| DLCTQA | NPFGFP | | | | | | PDBID:1BYL | | | | | | |
| PDBID:1KXW | | | | | | | RD | DD | DD | DQVV | RG | TNFRDA | PW |
| LDN | NF | TQA | NTDG | GILQ | SRWW | DGRTPGS | DPAG | | | | | | |
| NLCN | SSD | DGN | KGTD | RGC | | | PDBID:1EIF | | | | | | |
| PDBID:1LSL | | | | | | | KVG | DG | IFE | TSS | GD | LQTY | PEGI |

SVTC QMNG EA VTCG FGG DV NK

PDBID:1LYO

LDN NF TQA NTDG GILQ SRWW DGRTPGS

NLCN SSD GN KGTD IRGC

PDBID:1O4H

EEW KI NAENPRG SETTKG NAK LDSG SR

ADGLCHR

PDBID:1OAP

QQNN DLDK NPSY KEKPAVL

PDBID:1QTO

VD DRD GD TD RAV DYADTSG PA

DPAG

PDBID:1RMD

CQICE LAD TSCK CPSCR AQDC

PDBID:1SV9

SS GCY GWG NG LY

PDBID:1U3Y

DRT VTG KD RQF YKDVNI RKSDL

PDBID:1UIE

LDN NF TQA NTDG GILQ SRWW DGRTPGS

NLCN SSD DGN KGT RGC

PDBID:1UIF

LDN NF TQA NTDG GILQ SRWW DGRTPGS

NLCN SSD GN KGTD RGC

PDBID:1W2L

SIDG RLVGP FKGLYG EDG QPGAKV QGYPNV

PDBID:1WJX

FTGS DG NL PVDPRR LGKVE KG NERG

PDBID:1Y49

FCNAFTG

PDBID:1Y9T

FSIPNDL TTNK SKC AG VDGC NKIN DG

PDBID:1Z3P

NLTKDR CKNG TGSS YP GNPY

PDBID:2A7B

NG SNTNPS PKTF PAFNT NPQK KG NNF

PDBID:2AZB

LWKRP GG DTGA CG TPA

PDBID:2B1Y

RAGT KSPI AEGH

PDBID:2BFH

KNGG HPDG EKSDPHI RG VCAN KEDG KCVTDEC

ESNN RKYTSW KRTG GPGQKAI

EPG VG

PDBID:1FKV

KD GY TQA NNDS GLFQ NKIW DDQNPHS

NICN

PDBID:1GP3

NPSTG SEK CG NENCPPG GQTR KR DQ

CPSKSGL KQTC

PDBID:1GS3

GD NPFGQ SHNG NG DEHG

PDBID:1GSW

AFGSED DDG GDG IG DSPE DYQM ALSGD

PDBID:1HFX

NDLAG GY TQA SD GLFQ DKDF NICD

PDBID:1I9E

EG YPRQ VNG KSNS AS HWSDS FA

PDBID:1IGU

DPD EDDG DE LN AEG EDDG DPD

LKK DE LN PPAEG

PDBID:1IR8

LDN NF TQA NTDG GILQ SRWW DGRTPGS

NLCN SSD DGN KGTD RGC

PDBID:1JOI

AEC TDQM DKSCK PKNVMG GAG AAGT FPGHISMM

PDBID:1K58

DAKPQGR LTSPCKD ENKN REN WPPC NG HQSIFR

PDBID:1KJT

YPD APKARI LDKK SDL RAED NNV PTSA

EEDF

PDBID:1KXY

LNN NF TQA NTDG GILQ SRWW DGRTPGS

NLCN SSD DGN KGTD RGC

PDBID:1LAC

KLPD EGIH KPG NEDD NDKA SPVK LVP

VGQT

PDBID:1MKU

IPSS NN GCY NPYTN NN SSEN NLDKKN

PDBID:1NXT

PD MLPD AD KPF

PDBID:1O43

EEW KI NAENPRG SETTKG NAK LDSG SR

NS ADGLCHR

PDBID:1O44

EEW KI NAENPRG SETTKGA NAK LDSG SR

PDBID:2EYW

GAM DEED KKGD KPE DSEG VPYV PASA

PDBID:2GV2

DEKQQH CSND VKE

PDBID:2H2B

GSHM PGF FG NG MDNV SGK WRRT

PDBID:2H36

MEKV LGNY NG DDD EDN

PDBID:2IIY

AGDK ATWC YSN VDD CMP KG GAN

PDBID:2OXK

EK

PDBID:2POF

GG GN REPPPTA GWGPAGS LRGA GRHLS RR

TIPG DH

PDBID:2PKT

TPGS CIG DG TSNM CTD

PDBID:2QHE

NG TYY

PDBID:2R2Y

YL GT TDDS RTSG FPDD PQCPSG KAGS

EPKT

PDBID:2R34

TSI TPK TSI GERG

PDBID:2R36

TSI GERG PKT TSI LENY GERG

PDBID:2SAK

DSKG KPG TAYKE DPSA KNKK TEKG LSEHIKNP

PDBID:2WOR

RRDD NYLA DKNE

PDBID:2WOS

RRDD NYLA DKNE

PDBID:2X9C

KPSD KPSD

PDBID:2XKH

VN FGFKGV

PDBID:3AZ5

LDN RG NF TQA NTDG GILQ SRWW

DGRTPGS NLCN SSD NGMNAW KGT RGC

PDBID:3B84

GQYC GG GDG LALTSG

PDBID:3BYR

MDEG RGRA GP RGDT FPG

ADGLCHR

PDBID:1O4A

EEW KI NAENPRG SETTKGA NAK LDSG SR

ADGLCHR

PDBID:1O4I

EEW KI NAENPRG SETTKGA NAK LDSG SR

ADGLCHR

PDBID:1Q4V

GV SLYQL GERG GERG

PDBID:1QKF

SRR VG NGKQ

PDBID:1R75

FKNG MFSNI TEAR SKE YDS LGDT IG

RPL AEDERG

PDBID:1RAQ

LQCH DQGG NKVGP LHGIFG SGQAEGY IPGT

PDBID:1RBI

NLTKDR CKNG TGSS YP GNPY

PDBID:1RBW

DSST NLTKDR CKNGQTN TGSS YP GNPY

PDBID:1RNN

DSSTS NLTKDR CKNG TGSS YP GNPY

PDBID:1RNU

NLTKDR CKNG TGSS YP GNPY

PDBID:1RNW

DSST NLTKDR CKNG TGSS YP GNPY

PDBID:1RWJ

KGMT PK MKGV GMY CHTKLF KAGAKR

PDBID:1S7I

LAAV QGGR TKE

PDBID:1SFG

LA LDN RG NTDG GILQ SRWW DGRTPGS

NLCN SSDITA NRCKGT WIRGC

PDBID:1TDY

MDG GY TRA AGDR GIFQ SRWW DGKTPGA

NACH QDN DPQ QNR VQGC

PDBID:1TEN

TDT KDVPGD EDE KPDT GD

PDBID:1VAT

LDN NF TQA NTDG GILQ SRW DGRTPGS

LCN SSD DGN KGTD RGC

PDBID:1YEA

RC EEGG NKVGP LHG SGQVKGY IPGT

PDBID:3BZS

ST PTH KLGETPLP HKG

PDBID:3C12

DATG DANG DANG TAG SD TGL LANV

PDBID:3CTG

KE KTYC ELN TVP NG PVF

PDBID:3CYI

GAQG

PDBID:3HNU

NADG DG GG GAAE DCNG LNRI NG

PDBID:3KMJ

NDR SPLG RKS EKG TALEDY KN LGFGAIF

KDP TAGLK IKP AIG GDD RLTQ

PDBID:3KQ6

HSI HSI

PDBID:3LZ2

LDN NF THA NTDGS GILQ SRWW TPG

NLCN SSDI GGNGMNAW KGT IRGC

PDBID:3OSE

TWSM RF DARQD

PDBID:3RTO

TSI

PDBID:3S8S

GQIP RLNDNV HPRTR TS HL MG DIKG

TPQTV

PDBID:3WIT

NG EE PD GES KD GGS

PDBID:3WRP

SAA LKS

PDBID:3WX4

AG APGK

PDBID:4DP1

ADDGSLA PS PAG SEED AKG CSPH GM

PDBID:4F1A

TSI

PDBID:4IP1

KGK ADWC FSD ADT VTAN GLP DGQG

PDBID:4OV1

DQDKC APEL DDYG KGDG PAD CPEN

PDBID:4OZL

ADTP EKGD VRTG

PDBID:4PAZ

AE PA NPG VDK IKDM PEGA KINE

PDBID:1YMV

MPNMD DGAMSAL EAK PF

PDBID:1YVS

PDNY VAPGK NREGKLP SG NY SDW TDHYQT

PDBID:1Z3M

NLTKDR CKNG TGSS YP GNPY

PDBID:2BWK

DAKPKGR TSPC GAN REN GG RPPC NG

PDBID:2BWL

DAKPKGR TSPC GAN REN RPPC NG ESF

PDBID:2D4L

ATPGS PPN DNDY VNN SQG

PDBID:2DYQ

AR SDSL PL GRDPH LGRQS QPH

PDBID:2F3L

DV LIGE FSGK LTYA NA LTDS FSEA

LRGA GS LIGA LHGA LTNG FKGA LTNA

LTEA FDDA ITGA FSLA NPKTG

PDBID:2F4G

LDN NF TQA NTDG GILQ SRWW DGRTPGS

NLCN SSD NGMNAW KGTD RGC

PDBID:2FBO

NET EEG SSTRRD

PDBID:2FQL

SHG PAF PPNKQ PLS LNGE LRNG

PDBID:2M6D

CVYP

PDBID:2M6E

GC CV VY YP PW WC

PDBID:2M6G

GCPD

PDBID:2OLI

GCY NG LY

PDBID:2P6V

TPD

PDBID:2PW5

GD KG LL ETKVEKY AKK DKYG DG

KG

PDBID:2PYK

GD KG LL ETKVEKY AKK DKYG DG

KG

PDBID:2PZW

GD KG LL ETKVEKY AKK DKYG DG

CTAH GDSP

PDBID:4TS8

QDPES DPQLLG VKNP GQYQE RKTS

PDBID:4XXL

GVDK KDGP RGK FPDG INGK

PDBID:5B52

RLA

PDBID:5ER4

EGDKH SHFL DSDG

PDBID:5GSP

CD GS GSNS NYEG LSSG ENN HTGASGNN

PDBID:5LAZ

EKQQH CSND VKE

PDBID:5UES

DVEALGLH IKHP PPDH

PDBID:5YI1

ATNDERV

PDBID:6CED

CPH PAAG DVTQ CGDCG IQE LSCY GRYING

YIDL YYCQ

PDBID:6FGL

HDAAWPFL NPRLVSG IKNP EDDS

PDBID:6HOK

LDN NF TQA NTDG GILQ SRWW DGRTPGS

NLCN SSD DGNGMNAW KGTD IRGC

PDBID:155C

NEG KCKAC GKT NPDL GKN AXXXX

PDBID:1ANG

QDN DAKPQGR GLTSPCKD ENKN REN WPPC NG

PDBID:1AQT

AEQ SEG YPGH QHG PG SHG

PDBID:1BXY

PIGY RLQ PIGY RLQ AHL

PDBID:1CDP

FAGVL AADS GLTSK DQDK FKADA DSDG

PDBID:1CDT

KLIPIA PEGKN ASKKM SAL TDRC KLIPIA PEGK

ASKKM NVC SAL TDRC

PDBID:1DPY

GCY PG NPNIK QP DSAD ML STSC

PDBID:1DQ7

GDN PY PTPVP DGDN PY PTPV

KG

PDBID:2RKN

CG KENP YKNS PT TC

PDBID:2RNS

NLTKDR CKNG TGSS YP GNPY

PDBID:2VJW

AV EA DV DGPG TD SGA

PDBID:2VP7

SSSD GICT NDDQ EASC EAS CDTCM TE

GQVETIVS

PDBID:2XDY

GSKANEG HQ PRK TFGV STKG

PDBID:2XFE

VNG GG TG NG TGV NN GG

HCN

PDBID:2XMU

VPTI DAQA LTSK VPTI EDAQA LTSK

PDBID:2YXY

HPG EG GV LKG

PDBID:2Z7J

IYFA GHR DPNE YLDY AEKS APGQA HGR

PDBID:3AOD

SPN TG GPS QGDC SG TGGLGSGC HNNG

DQSN QQDR

PDBID:3ADY

NTLTIP PSVP PNSQ

PDBID:3BF2

LTYR DAAGA VTAVI RG

PDBID:3FZA

NP KTWC TVP GG

PDBID:3I7T

SR GP TD KIR LIAPGL

PDBID:3I7Y

DSSTS NLTKDR KP CKNG TGSS YP GNPY

PDBID:3IJU

LDN NF TQA NTDG GILQ SRW DGRTPGS

LCN SSD DGN KGTD RGC

PDBID:3JTE

CNSI MKMPKLS TPHM GHG KPV

PDBID:3MAZ

PAC GSDS IDI GQ LEKP RGNLR

PDBID:3NGR

| | | | | | | | | | | | | | |
|---|---|---|---|---|---|---|---|---|---|---|---|---|---|
| PDBID:1F1W | | | | | | | QSGP | KN | ATN | QPGT | DVGRNV | SVES | |
| EEW | KI | NPENPRG | SETTKGA | NAK | LDSG | SR | PDBID:3R5P | | | | | | |
| ADGLCHR | | | | | | | TFQK | GRKTG | GG | AEKN | KK | | |
| PDBID:1FLU | | | | | | | PDBID:3RH1 | | | | | | |
| LDN | NF | TQA | NTDG | GILQ | SRWW | DARTPGS | DSST | NLTKDR | CKNG | TGSS | YP | GNAY | |
| NLCN | SSD | DGNGMNAW | KGTD | RGC | | | PDBID:3RSK | | | | | | |
| PDBID:1FLW | | | | | | | DSST | NLTKDR | CANG | TGSS | YP | GNPY | |
| LDN | NF | TQA | NTDG | GILQ | SRWW | DGRTPAS | PDBID:3S3Y | | | | | | |
| NLCN | SSD | GNGMNAW | KGTD | RGC | | | GS | RTNG | LNSV | DG | GSS | TRAQ | DG |
| PDBID:1HQB | | | | | | | PDBID:3ZK0 | | | | | | |
| SQFG | PVSEF | WDT | | | | | SPGAE | SDSM | EA | DG | DR | PAH | KSGG |
| PDBID:1HSQ | | | | | | | KL | KQG | AH | | | | |
| AQREDEL | IKS | EGG | YGGK | | | | PDBID:4ET8 | | | | | | |
| PDBID:1I07 | | | | | | | LDN | NF | TQA | NTDG | GILQ | SRW | DGRTPGS |
| NSS | MKD | ASG | NNI | ARNSSE | MKD | NASG | LCN | SSD | DGNGMNA W | KGTD | RGC | | |
| NNI | | | | | | | PDBID:4FC1 | | | | | | |
| PDBID:1IR7 | | | | | | | PGDYA | | | | | | |
| LDN | NF | TQA | NTDG | GILQ | SRWW | DGRTPGS | PDBID:4FDX | | | | | | |
| NLCN | SSD | DGN | KGTD | RGC | | | GVEHF | AG | GVEHF | AG | | | |
| PDBID:1ISU | | | | | | | PDBID:4IP6 | | | | | | |
| NG | AQ | GASPTA | KV | PGDNQ | APGG | CDAF | KGK | ADWC | FSD | ADT | VTAN | GLP | DGQG |
| NG | AQ | SPTA | KV | PGDNQ | APGG | CDAF | PDBID:4NOZ | | | | | | |
| PDBID:1IZA | | | | | | | NI | KTEC | IEKN | SVP | NK | | |
| TSI | SLYQL | FVN | TSI | | | | PDBID:4WBF | | | | | | |
| PDBID:1LCJ | | | | | | | KER | ATN | APGT | SIDENA | | | |
| EPW | FKNL | APGNTHG | SESTAGS | QNQ | DNG | PR | PDBID:4YX5 | | | | | | |
| SDGLCTR | | | | | | | PTN | ANGV | ND | YA | VEG | | |
| PDBID:1LPI | | | | | | | PDBID:5FG4 | | | | | | |
| LDN | NF | TQA | NTDG | GILQ | SRWW | DGRTPGS | DTGNIFS | IKKP | AKDT | | | | |
| NLCN | SSD | DGN | KGTD | RGC | | | PDBID:5HBO | | | | | | |
| PDBID:1M4A | | | | | | | IPE | HE | APA | ED | | | |
| KKA | RPR | | | | | | PDBID:5MXY | | | | | | |
| PDBID:1MLI | | | | | | | GVDG | KFAD | PEM | FKEKL | | | |
| IAGH | HPSS | IAGH | HPSS | IAGH | HPSS | IAGH | PDBID:6BSY | | | | | | |
| HPSS | IAGH | HPSS | IAGH | HPSS | IAGH | HPSS | NPE | | | | | | |
| IAGH | HPSS | IAGH | HPSS | IAGH | HPSS | | PDBID:6I5A | | | | | | |
| PDBID:1NN7 | | | | | | | QFQT | QVTV | PDSD | RYN | DT | ENQA | GTG |
| SG | YPDTLLGS | PRHE | LIPE | | | | | | | | | | |

Table. S2. Amino acid sequences of β-turns of the 1000 protein samples.

**Note S3. Amino acid sequences of long β-strands with two adjacent RB residues locating at the middle segments of the β-strands.**

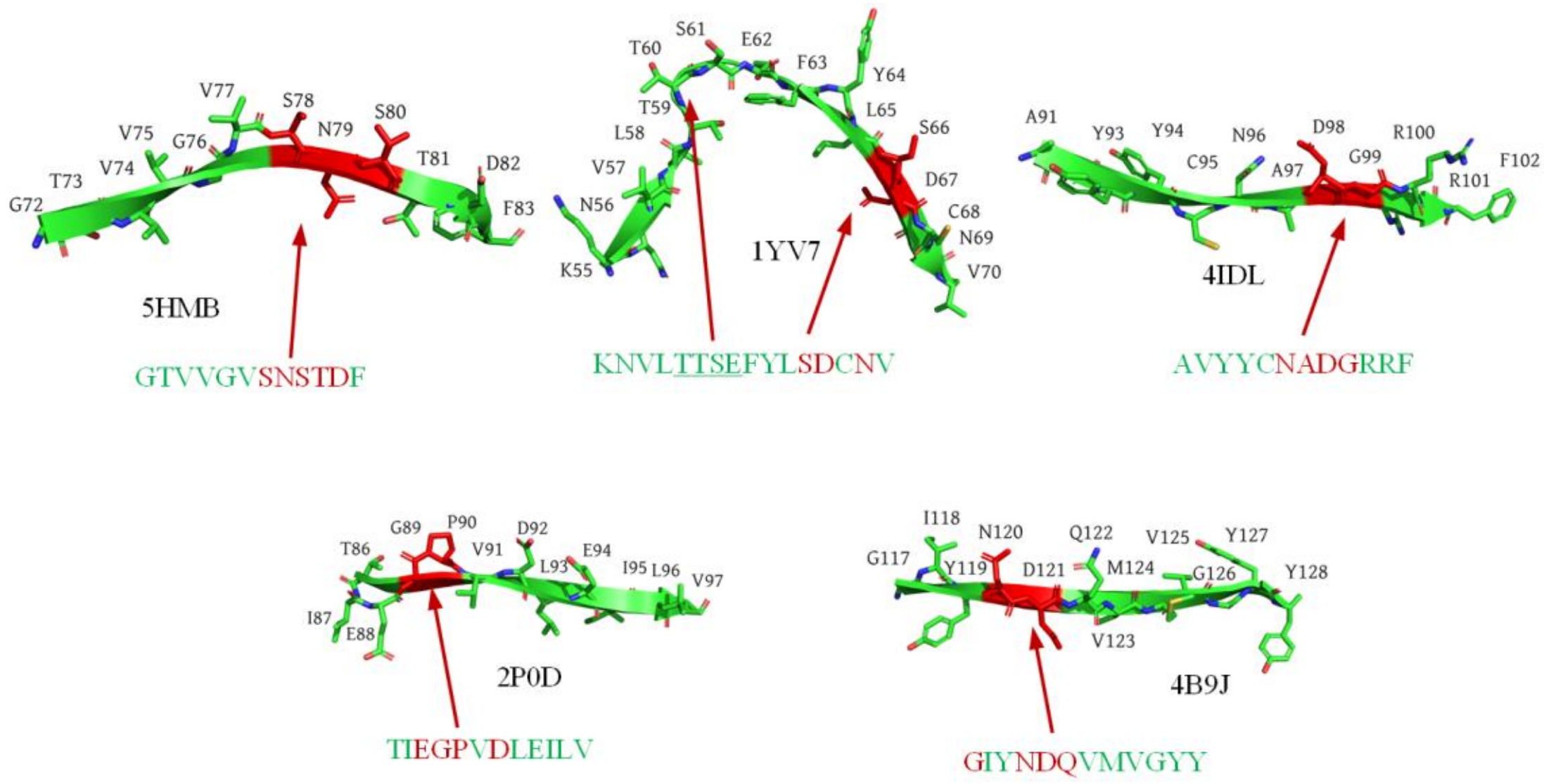

Figure. S1 Long β-strands haven't curved at their two RB residues in the amino acid sequences in the 1000 proteins.